\documentclass[12pt,a4paper]{article}
\usepackage{graphicx}
\usepackage{lscape}
\usepackage{natbib}
\setcitestyle{authoryear,open={(},close={)}}

\usepackage[titletoc,title]{appendix}
\usepackage{xcolor, colortbl}
\usepackage{float}
\usepackage{fullpage}%{geometry}
\usepackage{rotating}
\usepackage{amsfonts}
\usepackage{amsmath}
\usepackage{amssymb}
\usepackage{bbm}
\usepackage[strict]{changepage}
\usepackage{supertabular}
\usepackage{longtable}
\usepackage{tabularx}
\usepackage{fancyhdr}
\usepackage[shortlabels]{enumitem} %label enumerate 
\usepackage[font=scriptsize]{subfig}
\usepackage{array}

\usepackage[bookmarks=false, pagebackref=true, colorlinks=true, linkcolor=black, citecolor=black, breaklinks=true, urlcolor=black]{hyperref}
\definecolor{red_s}{rgb}{0.95 0.6 0.6}
\definecolor{green_s}{rgb}{0.72 0.85 0.69}
\definecolor{gray_s}{rgb}{0.9 0.9 0.9}

\input{prelim.sty}

\begin{document}

\begin{titlepage}

%\title{Measuring Macroeconomic Uncertainty:\\ A Data Revision Approach
%\title{Measuring Macroeconomic Uncertainty:\\ A Cross-Country Analysis
\title{Measuring Macroeconomic Uncertainty:\\ The Labor Channel of Uncertainty from a Cross-Country Perspective\thanks{We are grateful to Efrem Castelnuovo, Todd E. Clark, Jan Jacobs, Alexander Rathke, Michael Siegenthaler, Gregor von Schweinitz, and the participants of the Economics Research Seminar at Leipzig University for very helpful comments. We thank Tino Berger, Todd E. Clark, and Haroon Mumtaz for making available their global uncertainty estimates. We further thank Stefan Neuwirth and Roberto Golinelli for providing us with historic German and Italian GDP real time data. A previous version of the paper has been circulated under the title ``Measuring Macroeconomic Uncertainty: A Cross-Country Analysis''.}%
%\title{Measuring Macroeconomic Uncertainty From \\ Unpredictable Data Revisions
%\thanks{We are grateful to Alexander Rathke, Michael Siegenthaler, Gregor von Schweinitz, and the participants of the Economics Research Seminar at Leipzig University for very helpful comments. We thank Tino Berger, Todd E. Clark, and Haroon Mumtaz for making available their global uncertainty estimates. We further thank Stefan Neuwirth and Roberto Golinelli for providing us with historic German and Italian GDP real time data.}
}
%We are grateful to ...xxx..for helpful criticism and comments.

\author{Andreas Dibiasi\thanks{Center for Advanced Studies, EURAC Research, Drususallee 1, I-39100 Bozen, \href{mailto:andreas.dibiasi@eurac.edu}{andreas.dibiasi@eurac.edu}}\and Samad Sarferaz\thanks{KOF Swiss Economic Institute, ETH Zurich, Leonhardstrasse 21, CH-8092 Zurich, \href{mailto:sarferaz@kof.ethz.ch}{sarferaz@kof.ethz.ch}}}

	\date{\vspace{0.9cm}\today 
	%\\ \vspace{1cm} VERY PRELIMINARY - PLEASE DO NOT CIRCULATE
	}

\maketitle
\thispagestyle{empty}
\begin{abstract}\noindent
\noindent This paper constructs internationally consistent measures of macroeconomic uncertainty. Our econometric framework extracts uncertainty from revisions in data obtained from standardized national accounts. Applying our model to post-WWII real-time data, we estimate macroeconomic uncertainty for 39 countries. The cross-country dimension of our uncertainty data allows us to study the impact of uncertainty shocks under different employment protection legislation. Our empirical findings suggest that the effects of uncertainty shocks are stronger and more persistent in countries with low employment protection compared to countries with high employment protection. These empirical findings are in line with a theoretical model under varying firing cost. 

\noindent\\[0.1in]


\end{abstract}

\end{titlepage}
\smallskip

\linespread{1.4}
\selectfont
\section{Introduction}

\noindent The COVID-19 crisis has highlighted the relationship between uncertainty and economic fluctuations (e.g. \cite{altig2020economic}).\footnote{See \cite{Bloom2014} and \cite{castelnuovo2019domestic} for overview articles on the relationship between uncertainty.} % and economic fluctuations and \cite{altig2020economic} for the role of uncertainty during the COVID-19 pandemic.} 
This relationship usually differs across countries with conditions on the labor market possibly playing a decisive role.\footnote{The other prominent channels are the investment channel (see, for instance, \cite{Bloom2009} and \cite{bloom2018really}) and the financial channel (see, for instance, \cite{gilchrist2012credit}, \cite{LudvigsonMaNg2015}, and \cite{fernandez2020uncertainty})} However, while theoretically important, the empirical evidence on labor market specific transmission mechanisms of uncertainty shocks on the overall economy are so far scant. \\

\noindent This paper proceeds in two steps. First it constructs measures of macroeconomic uncertainty that are available for a large set of countries and that are defined as the conditional volatility of an unpredictable forecast as in \cite{Juradoetal2015}.\footnote{See also \cite{cascaldi2020certain} for a comprehensive overview of different types of uncertainty measures.} To obtain this goal, we draw on the macroeconomic data revisions literature, thereby treating statistical agencies' estimates of first releases of macroeconomic variables as forecasting exercises and their subsequent revisions as forecast errors.\footnote{See \cite{CroushoreStark2001} for the construction of real-time data sets and their relevance for macroeconomic research.} We extract the unpredictable part of data revisions by decomposing them into news -- the error from an unpredictable rational forecast -- and noise, which is defined as a classical errors-in-variables. Specifically, we follow the approach of \cite{JacobsvanNorden2011} in modeling data revisions with news and noise, enriching it with stochastic volatility components. Our measure of macroeconomic uncertainty is thus defined as the conditional volatility of the error corresponding to the unpredictable part of revisions in GDP growth. It is important to note that these estimates of macroeconomic uncertainty are consistent across OECD countries, given the nature of standardized national accounting procedures.\footnote{Statistical agencies in OECD countries follow similar national accounting standards. The data provided by the OECD database is based on the 2008 System of National Account. See also the website of the OECD for an overview of national legislation insuring the implementation of international accounting standards (\url{http://www.oecd.org/sdd/na/implementingthesystemofnationalaccount2008.htm}).} Note also that in constructing coherent macroeconomic variables statistical agencies take into account a plethora of series that include a huge amount of sensitive data partly only available to the statistical agency.\footnote{In Appendix \ref{sec:app__gdp}, we show that statistical agencies have valuable information regarding GDP growth that is not even exceeded by financial markets.} Thus, in contrast to the bottom-up approach of, e.g., \cite{Juradoetal2015}, we follow a top-down approach, where we partly outsource the information acquisition to the statistical agency.\\ %\footnote{xxx Referenzpaper checkenXX Statistical agencies in OECD countries follow similar national accounting standards. For instance, the data provided by the OECD database is based on the 2008 System of National Account.}   \\%\footnote{xx data on taxes, consumption, firm specifics??? cite literature}\\
%Further, the website of the OECD (\url{http://www.oecd.org/sdd/na/implementingthesystemofnationalaccount2008.htm}) presents an overview of national legislation insuring the implementation of international accounting standards.

\noindent We apply our procedure to a post WWII real-time dataset collected for 39 countries, deriving an international set of estimates of macroeconomic uncertainty. For the U.S., our measure pinpoints periods of highest uncertainty during the mid-1970s, beginning of 1980s, beginning of 2000s, and during the recent great financial crisis, which is qualitatively consistent with other measures of U.S. uncertainty. However, our measure already peaks in the mid-1970s, highlighting the turmoils during the 1970s. We also construct a global uncertainty measure by using a GDP weighted average of all country specific uncertainty indicators. According to our measure of global uncertainty, the period during the mid-1970s and the great financial crisis stands out in terms of uncertainty, which is in line with most of the measures of global uncertainty.\footnote{We compare our measure of uncertainty with the global uncertainty indicators presented in \cite{mumtaz2017common}, \cite{redl2018uncertainty}, \cite{carriero2019asssessing}  and \cite{berger2017global}.} We also perform a VAR analysis for the U.S. and the G7 countries. The impulse response functions computed for the U.S. are very similar to impulse responses from a VAR including the uncertainty indicator of \cite{Juradoetal2015}. These impulse response functions are qualitatively confirmed by the impulse responses estimated for an aggregate of the G7 countries.\\

\noindent In a second step, we use our newly created international set of indicators to investigate the role of labor adjustment costs in transmitting uncertainty shocks. We subdivide the countries into high employment protection legislation (EPL) countries and low employment protection legislation countries using the OECD Employment Protection Database. In an 8-variable VAR analysis that uses data from 1988Q1 to 2019Q4, we find that the degree of labor protection plays a crucial role in the propagation of uncertainty shocks. Uncertainty shocks affect the economy in countries with stricter employment protection legislation less than in countries with low labor protection standards. To learn more about the role played by EPL in the propagation mechanism of uncertainty shocks, we employ the theoretical model of \cite{bloom2018really}. Within their framework, our focus is on the effects of changes in firing costs, assuming that stricter employment protection legislation hinders firms to lay off employees and thus lead to higher firing costs. We first calibrate, solve, and simulate the model of \cite{bloom2018really} twice, once for an economy for low EPL and once for an economy with high EPL. We then use the two calibrated models to simulate the reaction of the economy to an imposed uncertainty shock. According to the theoretical model and in line with our empirical findings, an uncertainty shock has less deteriorating effects in an economy with high EPL than in one with low EPL.\\

%estimates to study the effects of macroeconomic uncertainty on the overall economy
\noindent There is a rapidly expanding literature that uses forecast error based procedures to estimate uncertainty as in, e.g.,  \cite{Juradoetal2015} and \cite{carriero2018measuring} who employ factor stochastic volatility models to provide uncertainty measures for the U.S. \cite{mumtaz2017common}, \cite{carriero2019asssessing}, \cite{berger2017global} estimate similar uncertainty indicators with a focus on extracting global uncertainty.\footnote{See also \cite{redl2018uncertainty} for an application of the \cite{Juradoetal2015} approach to multiple countries and \cite{caggiano2020} for global uncertainty estimates derived from hierarchical dynamic factor models.} We contribute to this literature by using real-time data on GDP growth and by computing forecast errors that incorporate information available to the economic agents up to and including at time $t$. The focus so far has only been on the last vintage of economic series at time $T$ thus incorporating information that goes beyond time $t$ with $t<T$.\footnote{\cite{rogers2019well} show that uncertainty indicators based on real-time data can considerable differ from their ex-post counterparts.} Limiting the construction of forecast errors to information that economic agents had about aggregate variables at time $t$ is in line with the above mentioned definition of macroeconomic uncertainty and might have important implications. %For instance, up to the first oil price shock in the early 1970s, economic agents might not have been aware of the importance of the oil price for the economy as the data prior to the oil price shock were uninformative about this relationship. Thus, forecast errors during the 1970s recessions are different if one trains a forecasting model with data up to that point in time or uses information that goes beyond the 1970s.
Moreover, the forecast error based uncertainty literature estimates uncertainty from models that allow for changes in the underlying conditional volatility only. Drifts in the parameters of a macroeconomic model can however also stem from changes in the structure of the economy, usually modelled with time-varying coefficients. In such a setup, a model that includes stochastic volatility components only is probably misspecified.\footnote{See \cite{cogleysargent2005} for a more detailed discussion of these kind of misspecifications within the context of VARs. See also the discussions in \cite{sims2001} and \cite{stock2001}.} Not including time-varying coefficients possibly attributes too much variation to the stochastic volatility components, thus exaggerating the fluctuations in the uncertainty measures. We tackle this issue by allowing the coefficients of our model to change over time.\\ 

\noindent The real-time forecast errors constructed with our approach are similar to those that are computed from the survey of professional forecasters as in \cite{josekkel2019}, \cite{clarkmccrackenmertens2020}, \cite{ozturk2018measuring}, and \cite{rossi2015}. While the forecast error in our approach is constructed from the unpredictable part of revisions of GDP growth, the forecast error computed from the survey of professional forecasters are focused on single releases of GDP growth and do not incorporate revisions in GDP growth. Data revisions can thus directly affect the uncertainty measures derived from survey of professional forecasters, whereas the uncertainty indicator estimated by our procedure takes these data revisions into account. %\footnote{\cite{josekkel2019}, for example, find data revisions to affect the relative size of major peaks in their uncertainty estimates.}  
Moreover, uncertainty estimates from survey of professional forecasters are restricted to a few countries and cannot be applied to a broad-based international setting, which is the focus of this paper. There is also a literature on international uncertainty indicators that are derived from textual data.\footnote{See \cite{baker2016measuring}, \cite{davis2016index}, \cite{hassan2020global}, and \cite{ahir2018world}  for a more detailed discussion.} While these methods have their merits especially when it comes to real-time tracking of uncertainty, they coincide only in special cases with forecast error based uncertainty measures. Moreover, comparing uncertainty indicators from textual analysis across countries requires not only mapping the semantic meaning of words into another language but also to consider aspects of intercultural communication in order to ensure a mutual meaning of the textual analysis.\footnote{See, for instance, \cite{harmsen2003cultures} and \cite{kwon2009assessing} for the impact of intercultural communication on mutual understanding.}\\

\noindent The remainder of the paper is structured as follows: In Section 2, we discuss the econometric framework and the estimation procedure. Section 3 discusses the construction of the real-time data set that serves as the basis for the uncertainty indicator. Section 4 and 5 present the uncertainty indicators and evaluate the macroeconomic relevance of uncertainty shocks. Section 6 examines the role of labor adjustment costs in the propagation of uncertainty shocks and Section 7 concludes.

\section{Econometric Framework}\label{sec:econometric}
In this section, we describe our econometric model, show how to derive direct measures of macroeconomic uncertainty, and how we cope with structural change. Finally, we briefly discuss our estimation procedure.\\

\noindent We follow the standard notation in the data revision literature where $y_{t}^{t+j}$ denotes an estimate published at time $t+j$ of some real-valued scalar variable $y$ at time $t$ for $t=1,...,T$ and $j=1,...,L$. According to \cite{JacobsvanNorden2011} the $j$th release of $y$ can be express as a function of its ``true'' value and a measurement error that is decomposed into a news and a noise term 
\begin{equation}
y_{t}^{t+j} = \tilde{y}_{t} + \nu_{t}^{t+j} + \zeta_{t}^{t+j}, 
\end{equation}
where $\tilde{y}_{t}$ represents the true value, $\nu_{t}^{t+j}$ the news component and $\zeta_{t}^{t+j}$ the noise component.\footnote{See \cite{KishorKoenig2012} for another framework that allows the estimation of both news and noise type measurement errors in data revisions. See \cite{JacobsvanNorden2011} for a more detailed discussion of the data revision literature.}
Within this setup the noise component is interpreted as classical errors-in-variables and the news component as a rational forecast error.\footnote{See \cite{MankiwRunkleShapiro1984}, \cite{MankiwShapiro1986} and \cite{deJong1987}, where measurement errors are described as news. See \cite{Sargent1989} for a statistical agency that estimates the ``true'' value making full use of available information and thus resulting into unpredictable revisions.} The main assumptions to distinguish between news and noise innovations are their correlation with the underlying true value of the variable. It is thus assumed that the news component carries information about the ``true'' value of the variable (i.e.  $E[\tilde{y}_{t},\nu_{t}^{t+j}]\neq 0$), whereas the noise components are independent of the ``true'' value of the variable (i.e., $E[ \tilde{y}_{t} ,\zeta_{t}^{t+j}]=0$), with $E[ \nu_{t}^{t+j} ,\zeta_{t}^{t+j}]=0$ for all $t$ and $j$.\\

\noindent To estimate those news and noise components, we build on the state space model developed by \cite{JacobsvanNorden2011}, which can be expressed as
%To estimate macroeconomic uncertainty from data revisions, we build on a modified version of the \cite{JacobsvanNorden2011} model which we enrich with stochastic volatility (SV) components.%\footnote{Note that in the \cite{JacobsvanNorden2011} framework there is an equation for all releases, whereas we only use two releases (first and second, first and third, etc.), because we are only interested in the stochastic volatility of the \emph{overall} news component, which includes all the intermediary news components.} 
%Consider the following state space model
\begin{align}
Y_{t}&=%
Z%
\alpha_{t}, \label{eq:measure} \\
\alpha_{t} &=%
\varphi + %
T %
\alpha_{t-1}  +%
R \eta_{t}, \label{eq:state}
\end{align}%%
where 
\begin{align}
Y_{t}=%
\begin{bmatrix} 
y_{t}^{t+1} \\ y_{t}^{t+L} 
\end{bmatrix},
Z&=% 
\begin{bmatrix}
1 & 1 & 0 &1 & 0 \\ 1 & 0 & 1 & 0  & 1
\end{bmatrix},
\alpha_{t}=%
\begin{bmatrix} 
\tilde{y}_{t} \\ \nu_{t}^{t+1}\\ \nu_{t}^{t+L}\\
\zeta_{t}^{t+1}\\ \zeta_{t}^{t+L}
\end{bmatrix},
\nonumber
\end{align}

\begin{align}
\varphi =%
\begin{bmatrix} 
c \\ 0 \\0 \\ 0 \\ 0
\end{bmatrix},
T=%
\begin{bmatrix}
\rho & 0 & 0 & 0 & 0\\ 0 & 0 & 0 & 0 & 0 \\ 0 & 0 & 0 & 0 & 0 \\ 0 & 0 & 0 & 0 & 0 \\ 0 & 0 & 0 & 0 & 0
\end{bmatrix},
R=%
\begin{bmatrix}
\sigma^{\nu 1}  & \sigma^{\nu L} & 0 & 0 \\ 0 & -\sigma^{\nu 1} & 0 & 0  \\ 0 & 0 & 0 & 0\\ 0 & 0 & \sigma^{\zeta1} & 0 \\ 0 & 0 & 0 & \sigma^{\zeta L}
\end{bmatrix},
\eta_{t} =%
\begin{bmatrix} 
\eta_{t}^{\nu 1}\\ \eta_{t}^{\nu L} \\ \eta_{t}^{\zeta1} \\ \eta_{t}^{\zeta L}
\end{bmatrix}\nonumber
\end{align}
with $\eta_t \sim N(0,I_4)$, where $c$ and $\rho$ are coefficients and $\sigma^i$ represents the standard deviation of $\eta_{t}^{i}$ for $i=\nu 1, \nu L,\zeta 1,\zeta L$.\footnote{See \cite{JSSvN2020} for a more general specification of this model.} %Now we show what to change in this setup to derive macroeconomic uncertainty. 

\subsection{Estimating Macroeconomic Uncertainty}%\subsection{Conditional Volatility of Unpredictable Data Revisions}
To obtain direct estimates of macroeconomic uncertainty, we define economic uncertainty similar to \cite{Juradoetal2015} as the conditional volatility of the unpredictable part of future values of the variable, i.e. in our case of subsequent releases of the variable. We thus treat the estimation procedure of early releases as a forecasting exercise. Within this context, the news components can then be seen as the unpredictable part of the forecast error. We obtain estimates of macroeconomic uncertainty by estimating changes in the variance of the news component. To do this we enrich the \cite{JacobsvanNorden2011} model with stochastic volatility components, modifying Equation (\ref{eq:state}) to
\begin{align}
\alpha_{t} &=%
\varphi + %
T %
\alpha_{t-1}  +%
R_t \eta_{t},  \label{eq:state2}
\end{align}%%
where 
\begin{equation*}
R_t=%
\begin{bmatrix}
\sigma_t^{\nu 1}  & \sigma_t^{\nu L}  & 0 & 0 \\ 0 & -\sigma_t^{\nu 1} & 0 & 0 \\ 0 & 0 &  0 & 0\\ 0 & 0 &  \sigma_t^{\zeta 1} & 0 \\ 0 & 0 & 0 & \sigma_t^{\zeta L}
\end{bmatrix}
\end{equation*}
with $\sigma_t^i =\exp(h_{t}^{i})^{1/2}$ for $i=\nu 1,\nu L,\zeta 1,\zeta L$ and $\alpha_{t}$, $\varphi$, $T$ and $\eta_t$ specified as in (\ref{eq:state}). The volatility components are modelled as latent variables whose logarithms are assumed to follow independent AR(1) processes:
\begin{equation}
 h_{t}^{i}  = \mu^i + \phi^i (h_{t-1}^{i} - \mu^i) + \tau^{i}\epsilon_{t}^{i},
\end{equation}
where $\mu^i$, $\phi^i$, $\tau^i$ are parameters, $\epsilon_{t}^{i} \sim N(0,1)$ and $i=\nu 1,\nu L,\zeta 1,\zeta L$.\footnote{See, e.g., \cite{KimShephardChib1998}.}\\

\noindent Our measure of macroeconomic uncertainty at time $t$ can thus be expressed as 
\begin{equation*}
U_{t} \equiv \sigma_t^{\nu 1} + \sigma_t^{\nu L}, 
\end{equation*}
which is a combination of the conditional volatility of the two rational forecast errors.

\subsection{Capturing Structural Change}
In dynamical systems of macroeconomic developments variations in the underlying volatility are intertwined with changes in the structure of the economy, i.e. time-varying coefficients. If drifts in the system are characterized by structural changes in the economy, a model that includes stochastic volatility components only is possibly misspecified. In such a setup, not allowing for time-varying coefficients would possibly attribute too much time-variation to the stochastic volatility components, exaggerating the time variation in the stochastic volatility components.\\

 %In this paper we tackle this issue by assuming that structural change can occur potentially at any point in time and that they change the nature of the economy itself. 
\noindent In our setup, we model structural changes as variations in the stochastic properties of the ``true'' value $\tilde{y}_{t}$.\footnote{In periodical intervals, statistical agencies adjust the definition of national accounts to account for structural change. Appendix \ref{appendix: benchmark} addresses the nature of these revisions in greater detail.}
%\footnote{\cite{clementsgalvao2013} extend the model of \cite{JacobsvanNorden2011}, allowing for non‐zero mean revisions to capture the impact of data revisions to aggregate national account estimates.} \\
\noindent We capture this change by allowing the coefficients in the equation of $\tilde{y}_{t}$ to vary over time, modifying Equation (\ref{eq:state2}) to:
\begin{align}
\alpha_{t} &=%
\varphi_t + %
T_t%
\alpha_{t-1}  +%
R_t \eta_{t}, \label{eq:state3}
\end{align}

\noindent where 
\begin{align}
\varphi_t =%
\begin{bmatrix} 
c_t \\ 0 \\0 \\ 0 \\ 0
\end{bmatrix},
T_t=%
\begin{bmatrix}
\rho_t & 0 & 0 & 0 & 0\\ 0 & 0 & 0 & 0 & 0 \\ 0 & 0 & 0 & 0 & 0 \\ 0 & 0 & 0 & 0 & 0 \\ 0 & 0 & 0 & 0 & 0
\end{bmatrix}
\end{align}
and 
  \begin{align*}
\begin{bmatrix} c_t \\ \rho_t\end{bmatrix} &= \begin{bmatrix} c_{t-1} \\ \rho_{t-1}\end{bmatrix} + \begin{bmatrix} \epsilon_t^{c} \\ \epsilon_t^{\rho},\end{bmatrix} 
\end{align*}

\noindent with $[\epsilon_t^{c} \ \epsilon_t^{\rho}]'  \sim N(0,V)$ and  $\alpha_{t}$, $R_t$ and $\eta_t$ specified as in (\ref{eq:state2}).\\

%\noindent To obtain direct measures of macroeconomic uncertainty we use Equation (\ref{eq:measure}) as observation and Equation (\ref{eq:state3})  as state equation. 

\subsection{Priors}
We use priors that are as diffuse as possible. The prior on $V$ is assumed to follow an Inverse Wishart distribution. The shape parameter is set to 3. The prior for the scale parameter is optimized according to the length of the series, in order for the AR(1) to cover the range of possible values. The prior for the variance of the stochastic volatilities is assumed to follow an Inverse Gamma distribution. %We believe that changes in the volatility of the true value of GDP as well as the idiosyncratic errors of the releases should occur in a smoother fashion than changes of the volatility of the news component. Hence, we adjust the priors on the variance of the stochastic volatilities accordingly. We thus set priors on the variance of $\sigma_{i}^{\tilde{y}}$, $\sigma_{i}^{\zeta 1}$ and $\sigma_{i}^{\zeta L}$ tighter than on $\sigma_{i}^{\nu}$. For countries with long time series such as the U.S. or U.K. this prior has little influence on the posterior. Countries with short time series such as Israel or Estonia profit from these priors in terms of convergence. We also specify very diffuse priors for the initial conditions of $\alpha_{t}$. We set the prior for the mean to zero and for the variance to 100. 
We set the priors on the variance of the stochastic volatilities as uninformative as possible.

\subsection{Estimation Procedure}\label{sec:estprod}
We obtain draws from the posterior of our model's parameters using Markov Chain Monte Carlo methods. More specifically, we use Gibbs sampling.\footnote{The Gibbs sampling procedure was programmed in Julia and is available upon request.} The Gibbs sampler consists of the following blocks: 
\begin{enumerate}
    \item Draw $\alpha_{t}$ conditional on $\varphi_{t}$, $T_{t}$, $R_{t}$ and data $Y_t$ using a forward filtering backward sampling as described in, e.g., \cite{CarterKohn1994}, 
    \item Draw $h_t^i$, $\mu^i$, $\phi^i$, $\tau^i$ for $i=\nu 1,\nu L,\zeta 1,\zeta L$ conditional on $\alpha_{t}$, $\varphi_{t}$, $T_{t}$ using the ancillarity-sufficiency interweaving approach proposed by \cite{kastner2014ancillarity},
    \item Draw $\varphi_{t}$ and $T_{t}$ conditional on $\alpha_{t}$ and $R_{t}$ using the simulation smoothing approach introduced by \cite{McCauslandMillerPelletier2011}. 
    \item Draw $V$ conditional on $\alpha_{t}$, $R_{t}$, $\varphi_{t}$ $T_{t}$ from an Inverse Wishart distribution.
\end{enumerate}

\noindent See Appendix \ref{sec:PosteriorSimulation} for a more detailed discussion of our estimation procedure.

\section{Real-Time Data}\label{sec:data}

%We use data revisions in macroeconomic aggregates to obtain measures of macroeconomic uncertainty for various OECD countries. Real-time data releases of macroeconomic aggregates are thereby the key ingredient to construct the uncertainty indicator. In our preferred specification, we base the indicator on nominal GDP. The reason being that nominal estimates are derived from source data, while quantity estimates are derived from nominal estimates relying on different methods---deflation, quantity extrapolation, or direct valuation---depending on the availability of source data.\footnote{\cite{fox2017concepts} detail the construction of national account for the United States.} \\

\noindent We use data revisions in real GDP growth for 39 countries to construct the uncertainty indicator. Particularly, we use the first year-over-year growth rate of real GDP as a forecast for the final year-over-year growth rate of real GDP that we define as the growth rate published after three years.\footnote{Comparing the first release of GDP and to the 12$^{th}$ release of GDP creates a publication lag of our measure of uncertainty of three years. In order to obtain more recent estimates of macroeconomic uncertainty, we continuously decrease the distance between the first and the last release at the current edge.} In order to obtain a comprehensive data set for various countries, we need to tap and combine several data sources. The largest part of our data is provided by the \textit{Original Release Data and Revisions Database}. The \textit{Original Release Data and Revisions Database} is part of OECD Main Economic Indicators database \citep{oecd2017mei} and represents the central data source of this project. The database provides different releases of macroeconomic aggregates for many countries. This study uses data from 39 countries. Table \ref{tab:country_overview} provides an overview of the countries included in our study, the data provider and the first available data point. 
%including Australia, Austria, Belgium, Brazil, Canada, Chile, Czech Republic, Denmark, Estonia, Finland, France, Germany, Great Britain, Hungary, Greece, Iceland, India, Indonesia, Ireland, Israel, Italy, Japan, Korea, Luxembourg, Mexico, Netherlands, New Zealand, Norway, Portugal, Russia, Slovakia, Slovenia, South Africa, Spain, Sweden, Switzerland, Turkey and the United States. 
Unfortunately, the \textit{Original Release Data and Revisions Database} provides releases of macroeconomic variables only since 1999. For data prior to 1999, we need to rely on other data sources. We primarily use the data made available by the Federal Reserve Bank of Dallas for releases prior to 1999. \cite{fernandez2011real} collect real-time data for various economies including those that we use in this study. The authors assemble the dataset from original quarterly releases of different macroeconomic aggregates from 1962 to 1998. We currently use these data for all countries except the U.S., Germany, Italy, Australia and New Zealand. For the U.S., we use data provided by the Federal Reserve of Philadelphia as they provide more exhaustive data compared to the data provided by Federal Reserve of Dallas. For Germany we rely on data provided by \cite{boysen2012impact} and for Australia we use data provided by the Australian Real-Time Macroeconomic Database \citep{lee2012australian}. For New Zealand we use data provided by the Reserve Bank of New Zealand \citep{sleeman2006analysis} and for Italy, we also use data releases of ISTAT that were kindly provided by \cite{golinelli2008real}.\footnote{Appendix \ref{sec:app__real_time_data} describes these various data sources and outlines the construction of our data base in more detail. We will provide the final real-time dataset as well as the code to construct it upon request.}\\

\begin{table}[!htbp]
 \caption{Data Source of Uncertainty Indicators}\vspace{-0.75cm}
    \begin{center}
    \scalebox{0.68}{
    \begin{tabular}{lccc}
        Countries       & MEI Code  & Datasource prior to 1999  & Available since\\
    \rowcolor{green_s}  Australia       & AUS       & Real-Time Macroeconomic Database (U. Melbourne)   & 1967 Q3\\
     \rowcolor{gray_s}    Austria         & AUT       & FED Dallas                & 1999 Q3\\
      \rowcolor{gray_s}  Belgium         & BEL       & no data                   & 1999 Q1\\
      \rowcolor{gray_s}  Brazil          & BRA       & no data                   & 2000 Q2\\
     \rowcolor{green_s}   Canada          & CAN       & FED Dallas \& Bank of Canada                 & 1961 Q3\\
     \rowcolor{red_s}   Chile          & CHL       & no data                 & 2010 Q1\\
     \rowcolor{gray_s}   Czech Republic  & CZE       & no data                 & 1999 Q1\\
     \rowcolor{gray_s}   Denmark         & DNK       & FED Dallas                & 1993 Q2\\
     \rowcolor{red_s}   Estonia         & EST       & no data                 & 2010 Q3\\
      \rowcolor{gray_s}  Finland         & FIN       & FED Dallas                & 1993 Q4\\
    \rowcolor{green_s}    France          & FRA       & FED Dallas                & 1987 Q2\\
    \rowcolor{green_s}    Germany         & DEU       & FED Dallas \& \cite{boysen2012impact}                 & 1964 Q4\\
    \rowcolor{green_s}    Great Britain   & GBR       & FED Dallas                & 1964 Q4\\
    \rowcolor{red_s}    Hungary       & HUN       & no data                & 2002 Q2\\
     \rowcolor{red_s}    Greece          & GRC       & no data                   & 2003 Q4\\
      \rowcolor{red_s}   Iceland         & ISL       & no data                   & 2002 Q4\\
      \rowcolor{red_s}   India           & IND       & no data                   & 2005 Q4\\
      \rowcolor{red_s}   Indonesia       & IDN       & no data                   & 2005 Q4\\
      \rowcolor{red_s}   Ireland         & IRL       & no data                   & 2002 Q2\\
      \rowcolor{red_s}  Israel          & ISR       & no data                   & 2010 Q2\\
    \rowcolor{green_s}    Italy           & ITA       & FED Dallas \& \cite{golinelli2008real} & 1974 Q3\\
    \rowcolor{green_s}    Japan           & JPN       & FED Dallas                & 1964 Q3\\
    \rowcolor{gray_s}    Korea           & KOR       & FED Dallas                & 1996 Q4\\
      \rowcolor{red_s}   Luxembourg      & LUX       & no data                   & 2004 Q4\\
    \rowcolor{gray_s}    Mexico          & MEX       & FED Dallas                & 1994 Q2\\
     \rowcolor{gray_s}   Netherlands     & NLD       & FED Dallas                & 1993 Q3\\
      \rowcolor{gray_s}  New Zealand     & NZL       & Reserve Bank of New Zealand                & 1994 Q4\\
      \rowcolor{gray_s}  Norway          & NOR       & FED Dallas                & 1993 Q3\\
      \rowcolor{red_s}  Poland        & POL       & no data                & 2002 Q2\\
      \rowcolor{gray_s}  Portugal        & PRT       & FED Dallas                & 1992 Q3\\
      \rowcolor{gray_s}  Russia          & RUS       & no data                   & 1999 Q3\\
      \rowcolor{gray_s}   Slovakia    & SVK       & no data                   & 2000 Q3\\
      \rowcolor{red_s}   Slovenia    & SVN       & no data                   & 2010 Q1\\
      \rowcolor{red_s}   South Africa    & RUS       & no data                   & 2001 Q4\\
      \rowcolor{gray_s}  Spain           & ESP       & FED Dallas                & 1993 Q1\\
     \rowcolor{green_s}   Sweden          & SWE       & FED Dallas                & 1989 Q4\\
     \rowcolor{green_s}   Switzerland     & CHE       & FED Dallas                & 1987 Q2\\
     \rowcolor{gray_s}   Turkey          & TUR       & FED Dallas                & 1993 Q1\\
      \rowcolor{green_s}  USA             & USA       & FED Philadelphia          & 1961 Q4 
    \end{tabular}
    }
    \end{center}
        \footnotesize{Notes: Central data source is the \textit{Original Release Data and Revisions Database}. The column \textit{Countries} depicts the country and \textit{MEI Code} the country code from the \textit{Original Release Data and Revisions Database}. The column \textit{Datasource prior to 1999} describes the data source of releases prior to 1999. The column \textit{Available since} states beginning of a country's uncertainty indicator. Rows in green highlight countries with data available prior to 1990Q1, rows in gray depict countries with data from 1990Q1 to 2000Q4 and rows in red represent countries with data available only from 2001Q1 onward.}
    \label{tab:country_overview}
\end{table}

\newpage

\begin{figure}[!htbp]
    \centering
    \includegraphics[scale=0.8]{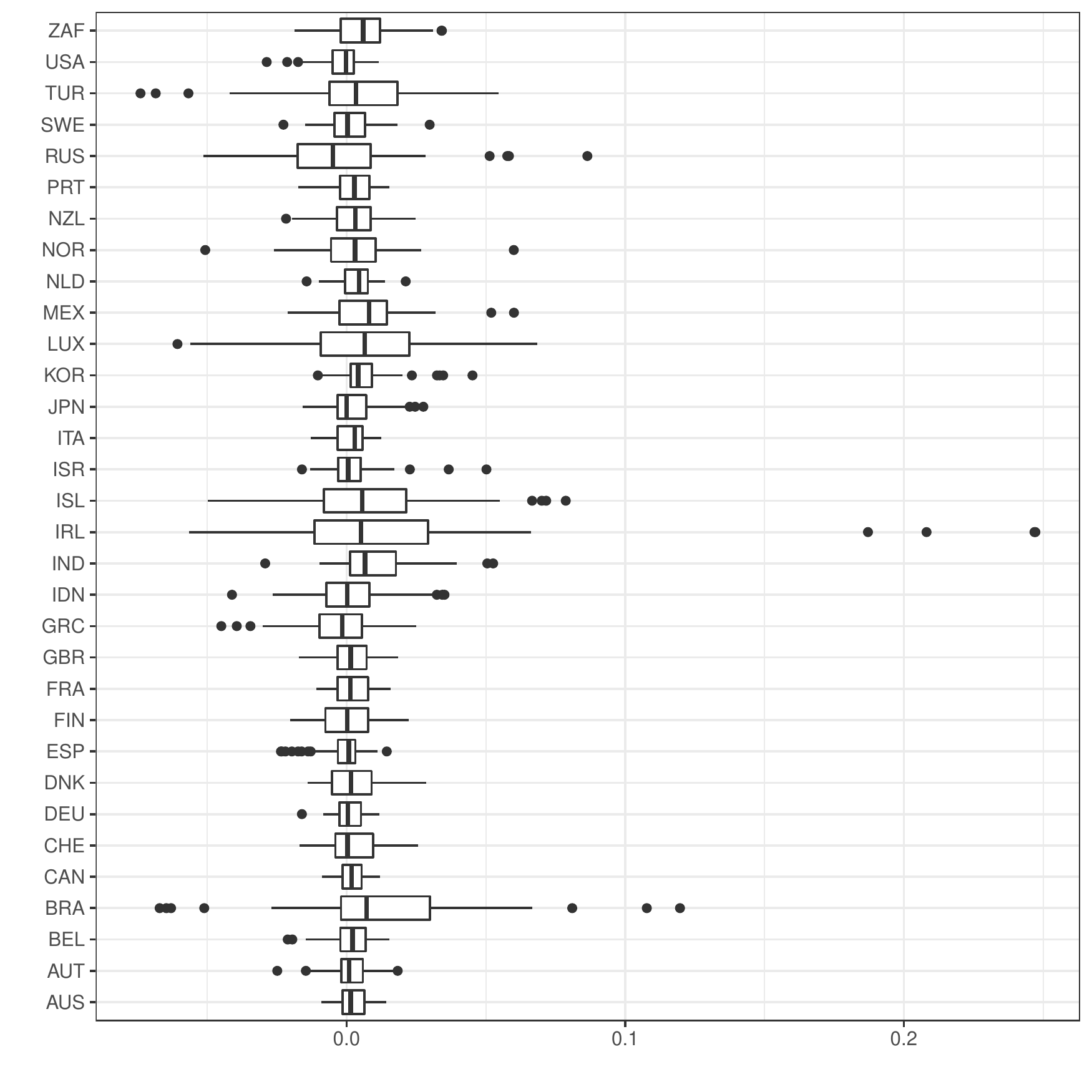}
    \caption{Boxplot of the 10th revision of real GDP growth for the periods from 2000 Q1 onward.}
    \label{fig:diverge_boxplot}
\end{figure}

\noindent While data availability fluctuates a lot between countries, we have surprisingly long time series for many countries. For 10 countries we have real-time data for more than 30 years (green shaded countries) and for another 16 countries we have more than 20 years of data (gray shaded countries). For four countries only, we have less than 10 years data. Besides data availability, also the average revisions change heavily between single countries. While large countries in terms of GDP, such as the U.S., France, Germany, Canada and Australia tend to have small revisions, smaller countries, including Ireland, Island and Luxembourg, appear to have much larger revisions. Figure \ref{fig:diverge_boxplot} visualizes the distribution of the 10th revision of year-over-year growth rates of real GDP for different countries. 

\noindent Most countries reveal a statistical significant upward revision of their growth rates over time. The 10th release of GDP growth tends on average to be larger than the first release. Only four countries including the U.S., Russia, Greece and Spain, report on average a lower growth rate at the 10th release than on the first release.\footnote{See Figure \ref{fig:diverge_bars} in Appendix \ref{sec:app__real_time_data} for a better overview of average revisions.}

\section{Estimates of Macroeconomic Uncertainty}\label{sec:macrouncertainty}
Using the econometric framework outlined in Section \ref{sec:econometric} and the real-time dataset described in Section \ref{sec:data}, we obtain estimates of macroeconomic uncertainty for 39 countries.\footnote{We provide all uncertainty indicators on our website.} We now present and discuss the resulting uncertainty measures. Thereby, we focus on uncertainty in the United States and global uncertainty.\\

\begin{figure}[!htbp]
    \centering
    \includegraphics[scale=0.80]{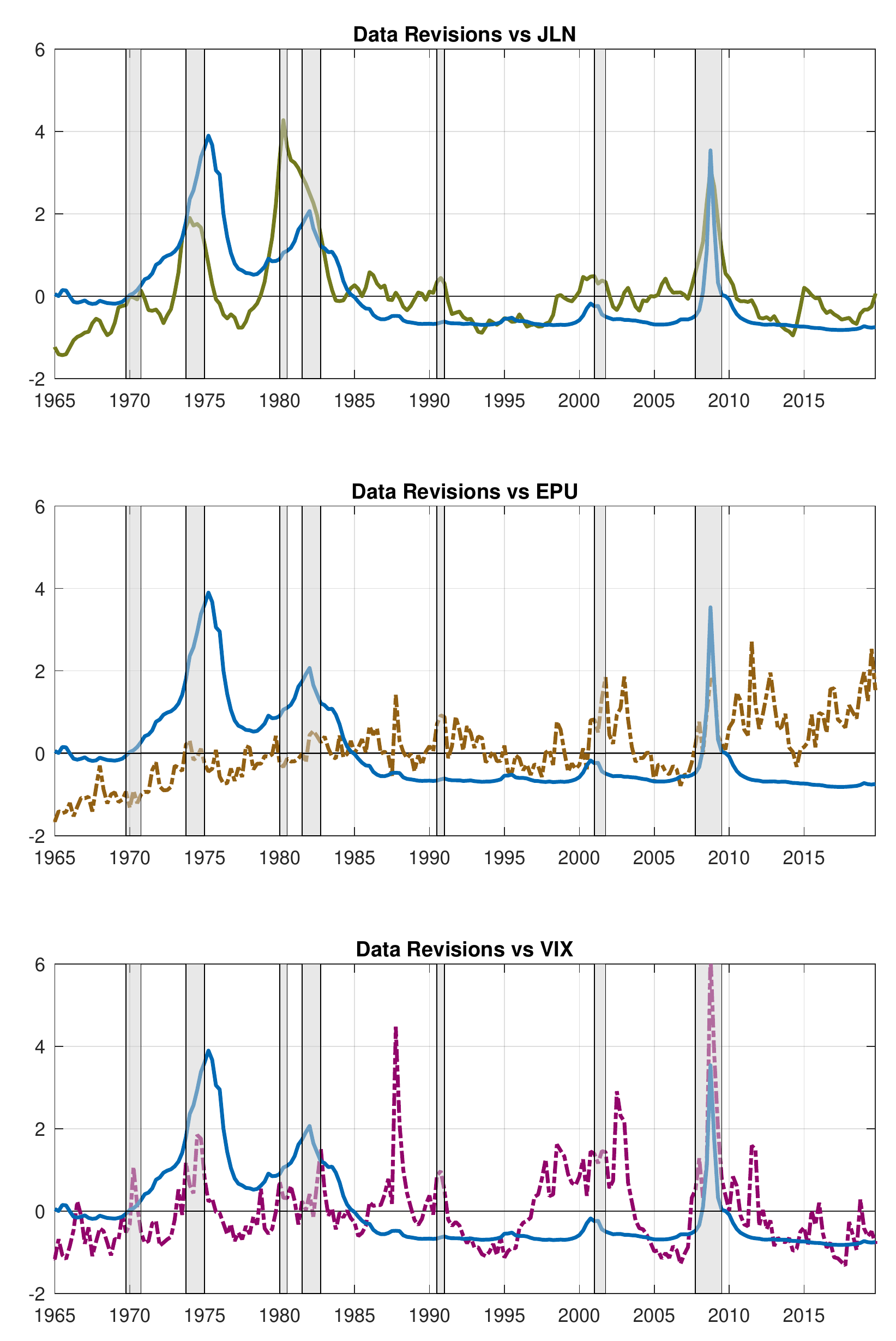}
    \caption{Uncertainty United States: Macroeconomic Uncertainty \newline \scriptsize{Notes: This figure compares different uncertainty indicators for the United States form 1960Q1 to 2019Q4. In the first pane, the green solid line displays the indicator for macroeconomic uncertainty (quarterly averages, horizon 12, MacroFinanceRealUncertainty\_202008\_update) developed by \cite{Juradoetal2015} and the blue solid line shows the newly proposed measure of macroeconomic uncertainty. In the second pane, the dashed ochre line shows quarterly average of the Economic Policy Indicator proposed by \cite{baker2016measuring}. The last pane compares the VIX (realisied volatility before 1989) to the new uncertainty measure. All indicators are demeaned and normalized to unit variance.}}
    \label{fig:gdp_vs_jln}
\end{figure}

\noindent Our methodical framework provides macroeconomic uncertainty estimates for the United States that are similar to existing uncertainty measures. Figure \ref{fig:gdp_vs_jln} presents our revision-based uncertainty measure for the U.S. (blue solid line) and compares it to existing proxies found in the literature. These alternative measures include the macroeconomic uncertainty indicator proposed by \cite{Juradoetal2015} (green solid line), the economic policy uncertainty index developed by \cite{baker2016measuring} (ochre dashed line) and the VIX (purple dashed line), a popular uncertainty indicator that reflects market's expectation of volatility implied by the S\&P 500 index options.\\

\noindent Our data revision based indicator reaches its highest levels during the recession in the 70s that was characterized by the first oil price shock and the collapse of the Bretton Woods system and marked the end to the overall Post-World War II economic expansion. The Great Recession of 2008 represents the second highest peak of our uncertainty measure. Further identified times of heighten uncertainty are during the '82 recession and, to a much lesser extent, at the beginning of the 2000s, during the dotcom bubble burst. Overall, our indicator resembles most the uncertainty measure proposed by \cite{Juradoetal2015}(henceforth JLN). However, while JLN peaks during the Iran Revolution in 1979, the revision based indicator reaches its highest levels during the 70s recession. Compared to other uncertainty measures, our uncertainty estimate for the U.S. displays a significantly lower volatility and indicates only a few mayor uncertainty shocks during 1965 and 2019. For instance, while both the EPU and the VIX peak in 1987 as a result of the Black Monday, the revision based indicators hardly blinks. Similar, after the Great Recession of 2008, the EPU reaches all-time-high level of economic policy uncertainty. However, the economic policy uncertainty does not translate into macroeconomic uncertainty as both the revision based indicator as well as JLN return to very low levels after the recession. \\

\noindent Although we obtain uncertainty estimates for 39 countries, for the sake of brevity, we abstain from discussing all countries in the main text and refer the reader to Appendix \ref{sec:app__unc} for a presentation of all uncertainty indicators.\footnote{Figure \ref{fig:allUnc1} and Figure \ref{fig:allUnc2} in the Appendix present the uncertainty estimates of all countries.} Instead, we use the comprehensive number of uncertainty indicators to examine uncertainty on a global level. We construct a global uncertainty measure as the weighted mean of single country indicators. We achieve this by first standardizing the uncertainty indicator of each country and then computing the weighted average by weighing each country according to its real GDP.\footnote{The global uncertainty indicator is based on an unbalanced sample. That is, countries' uncertainty estimates are considered according to their availability.} The countries included in the construction of the global uncertainty indicator account for approximately 50\% of world GDP during the first half of the sample. During the second half of the sample, the included countries account for more than 75\% of world GDP. Figure \ref{fig:allUnc2} presents the global uncertainty indicators. From the 1960s to today, our estimates suggest two large global uncertainty shocks. The first occurred during the oil price shock in the 70s, the second global uncertainty shock was experienced during the Great Recession in 2008. The only other notable increase in global uncertainty occurred after the second oil price shock and the subsequent early 80s recession. \\

\begin{figure}[h]
    \centering
    \includegraphics[scale=0.9]{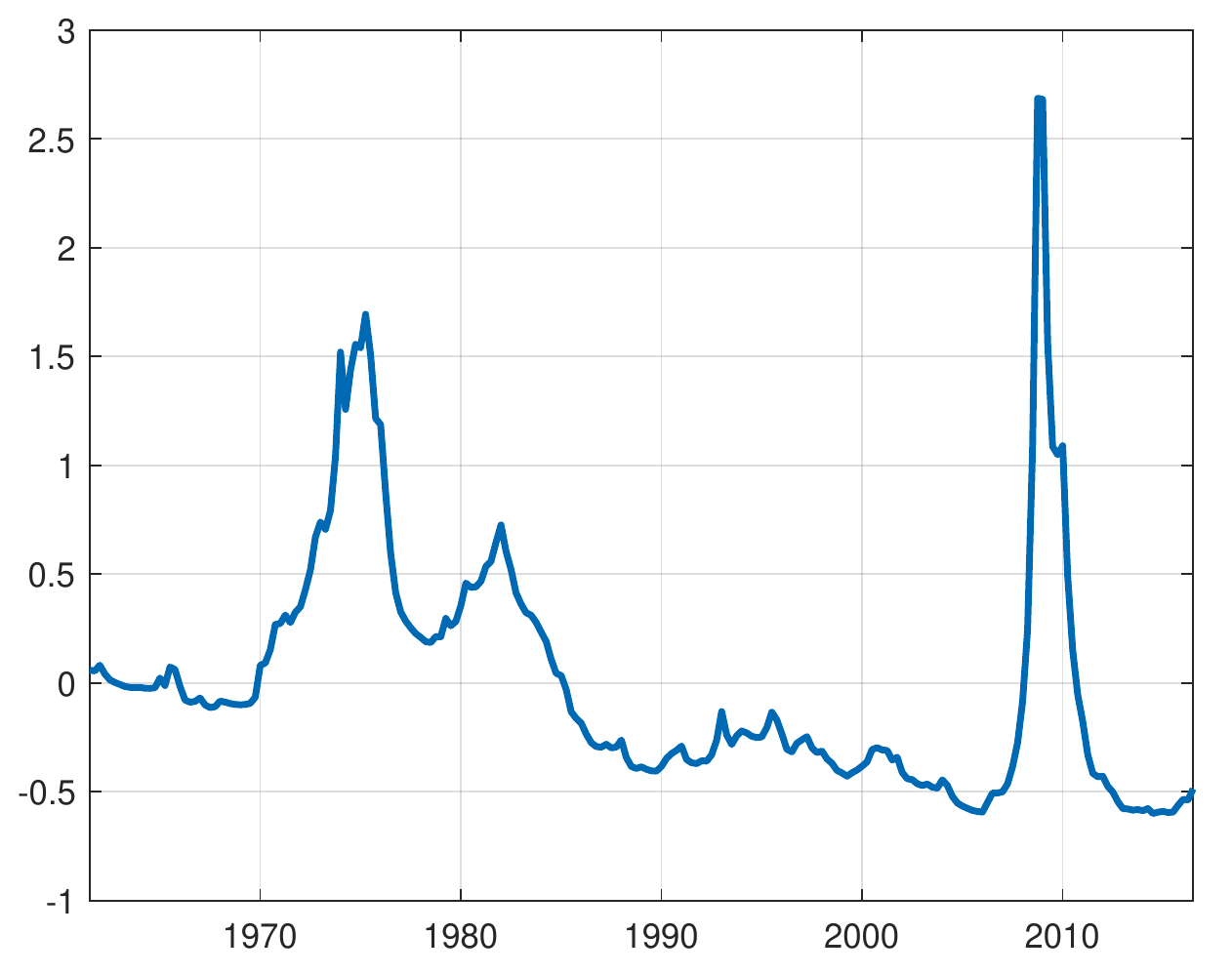}
    \caption{Global Uncertainty \newline \scriptsize{\newline
						Notes: This figure shows the weighted average of countries' normalized uncertainty measures. We weight single countries according to their real GDP. While the included countries represent approximately half of world GDP during the first half of the sample, the economies included in the second half represent around 75\% of total GDP.}
		}\label{fig:allUnc2}
\end{figure}

\begin{figure}[!htbp]
    \centering
    \includegraphics[scale=0.80]{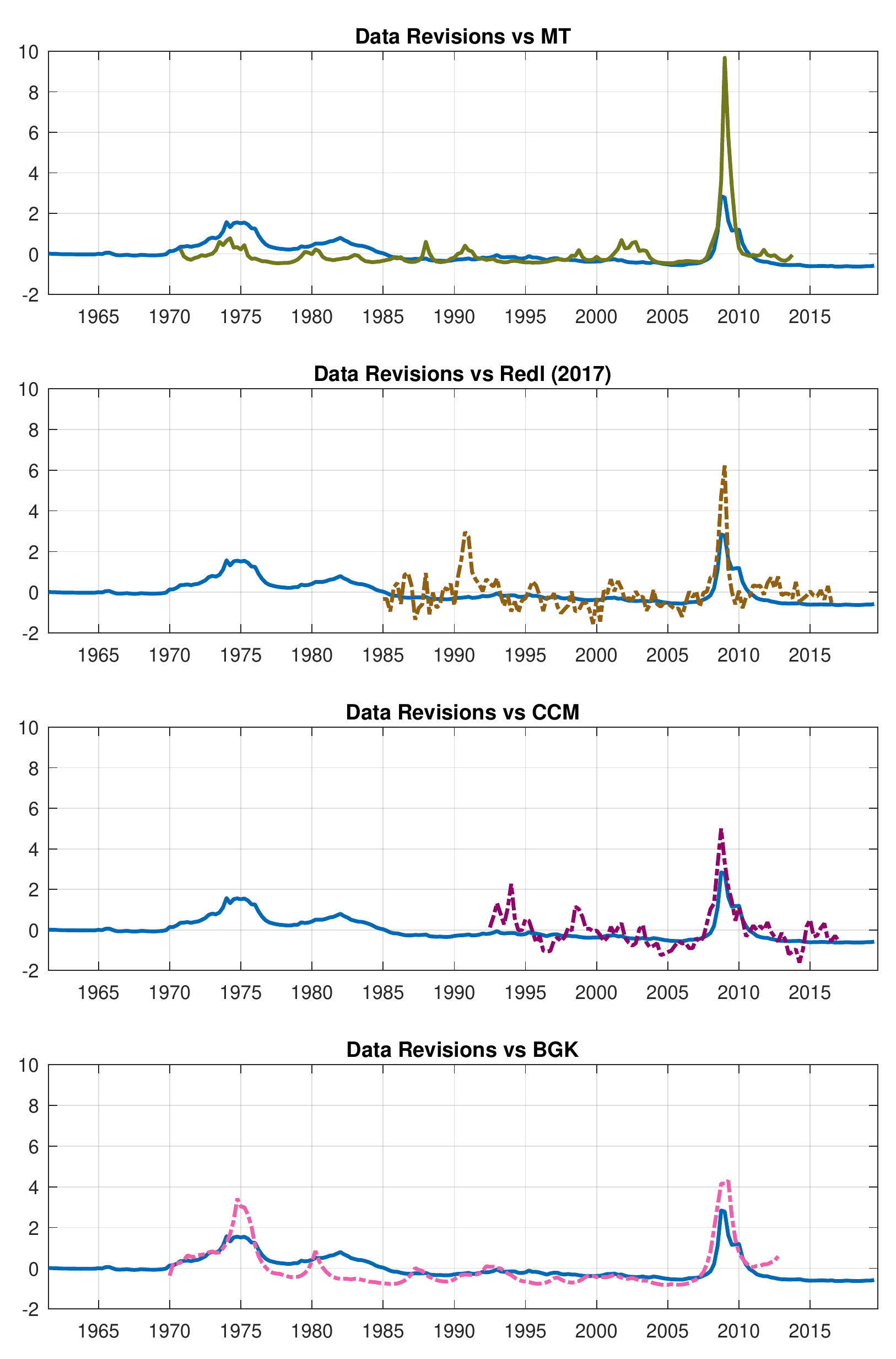}
    \caption{Global Macroeconomic Uncertainty \newline \scriptsize{Notes: This figure compares different global macroeconomic uncertainty indicators. In the first pane, the green solid line displays the indicator for global macroeconomic uncertainty by \cite{mumtaz2017common}  and the blue solid line shows the data revision based indicator of global macroeconomic uncertainty. In the second pane, the dashed ochre line shows quarterly averages of the global macroeconomic uncertainty measure proposed by \cite{redl2018uncertainty}. The third pane compares the global uncertainty indicator by \cite{carriero2019asssessing} to the new uncertainty measure. The last pane display the indicator proposed by \cite{berger2017global} together with the revision based indicator. All indicators are demeaned and normalized to unit variance.}}
    \label{fig:globUncComparision}
\end{figure}

\noindent Recently, various papers started to measure and study uncertainty on a global dimension.\footnote{See \cite{castelnuovo2019domestic} for a recent review on the literature focusing on global uncertainty.} While several papers propose measures of global economic policy uncertainty, world risk and global financial uncertainty, studies that attempt to measure global macroeconomic uncertainty are limited: \cite{redl2017impact} constructs a JLN based global uncertainty measure that uses global macro and financial data from emerging and advanced economies. \cite{mumtaz2017common} (henceforth MT) use a factor model with stochastic volatility to decompose the time-varying variance of macroeconomic and financial variables of eleven OECD countries\footnote{These countries include United States, United Kingdom, Canada, Germany, France, Spain, Italy, Netherlands, Sweden, Japan and Australia.} into contributions from country-specific uncertainty and uncertainty common to all countries. \cite{carriero2019asssessing} (henceforth CCM) estimate a large, heteroskedastic VAR on 19 industrialized economies to obtain estimates of global uncertainty. \cite{berger2017global} (henceforth BGK) use a dynamic factor model with stochastic volatility to identify the common component of macroeconomic uncertainty from 20 OECD countries. Figure \ref{fig:globUncComparision} compares these indicators to our data-revision based indicator. While the data revision based indicator matches the other indicators remarkably well, two differences stand out. First, similar to the indicator of the U.S., the data revision based indicator is the least volatile of all indicators. Second, while the other four indicators peak during the great recession of 2008 at 4 standard deviations or higher above their mean, the data revision indicator peaks at 2.5 standard deviations.

\section{Cross-Country Impact of Uncertainty Shocks}
Following the existing  empirical  research  on  uncertainty, we use a VAR analysis to study the dynamic relationships between macroeconomic activity and uncertainty. We consider the following VAR model:

\begin{equation}
x_t = c_b + B_1 x_{t-1} + B_2 x_{t-2} + ...+ B_p x_{t-p} + u_t,  \label{eq: reducedform}
\end{equation}
where $x_t$ is a $n \times 1$ vector containing all $n$ endogenous
variables, $t=1,...,T$ denotes time, $c_b$ is a $n \times 1$ vector of constants, $B_i$ for $i=1,...,p$
are $n \times n$ parameter matrices and $u_t$ is the $n \times 1$ one-step
ahead prediction error with $u_t\sim N(0,\Sigma)$, where $\Sigma$ is the $n \times n$ variance-covariance matrix. The prediction error $u_t$ can be written as a linear combination of structural innovations $u_t = A \epsilon_t$ with $\epsilon_t \sim N(0,I_n)$, where $I_n$ is an $(n \times n)$ identity
matrix and where $A$ is a non-singular parameter matrix.\\

\noindent We choose a recursive identification scheme and a VAR similar to the one proposed in  \cite{basu2017uncertainty}, augmenting their VAR setup with a stock market index. Similar to \cite{Bloom2009}, we include the stock-market level as the first variable in the VAR. We order uncertainty last to make sure that the impact of all other shocks is already considered for when evaluating the impact of uncertainty on the economy. The ordering of our VAR is as follows:\footnote{We have also experimented with the set-up of \cite{Bloom2009} by ordering uncertainty second right after the stock market variable. This reordering did not change the main results of this paper.}

\begin{align}
%VAR-8%
\begin{bmatrix}
\textit{stock market} \\ 
\textit{policy rate} \\
CPI \\
employment \\
investment \\
consumption \\
GDP \\
uncertainty 
\end{bmatrix}.
\end{align}\\
 We estimate the model using Bayesian methods, specifying diffuse priors.\footnote{We consulted the Bayesian information criterion and the Akaike information criterion for choosing a lag length. For the different countries and the different criterion, the suggested lag-lengths varied  from $p=1$ to $p=3$. We set the length to $p=2$ throughout this paper. The principal findings of the paper do not change when using lag length $p=3$ instead.} Similar to \cite{Juradoetal2015}, we use the posterior mean of our uncertainty indicator discussed in the Section \ref{sec:macrouncertainty} as a measure for macroeconomic uncertainty in our VAR. However, as a further robustness check, we have also estimated the VAR taking into account the uncertainty surrounding our uncertainty indicator. To incorporate the whole posterior distribution instead of just the posterior mean, we extend the algorithm in Section \ref{sec:estprod} by one further block. In this additional step, we obtain a draw for the VAR parameters from a Normal-Inverse Wishart distribution, conditional a draw simulated from the posterior of the uncertainty indicator. The resulting posterior distributions are summarized in Figures \ref{fig:irf_hl_genreg} and \ref{fig:irf_all_genreg} in Appendix \ref{sec:app__rob}.\\
\subsection{Impact of Uncertainty Shock in the U.S.}
%We use the described VAR to investigate the dynamic responses of key macro variables to innovations in our uncertainty measures and compare them to the responses to innovations in the macroeconomic uncertainty index by \cite{Juradoetal2015}. In line with the literature, we will refer to innovations to the uncertainty measures as ``shocks''. Figure \ref{fig:irf_us} presents the responses of GDP, investment, employment and consumption to a uncertainty shock in the United States. A one standard deviation increase of our uncertainty variable has an adverse and enduring effect on the economy (blue dashed line). The  90\%  standard-error  bands  (gray area) around the impulse responses highlight that drop is statistically significant. Thereby, the drop lasts for about two years, with output declining by around 0.4\%, investment by  1\% and employment by about 0.5\%. For all variables, the recovery from the shock takes up to 10 years and more. While these effects seems somewhat strong and long-lasting, they are very similar to the uncertainty measure of \cite{Juradoetal2015} (green line). 

We employ the described VAR to investigate the impulse responses functions of key macro variables to uncertainty shocks that are derived using our data revision based uncertainty measure. To validate our uncertainty measure, we compare the impulse responses obtained using the data revisions based uncertainty measure to impulse response functions that are computed with the macroeconomic uncertainty index by \cite{Juradoetal2015}. The estimation sample spans the period 1982Q1--2019Q4.\footnote{Due to irregularities in the revision scheme of U.S. GDP during the 1970s, we start the VAR analysis in 1982Q1 and not earlier. See Table 2 in \cite{fernandez2011real} for a more detailed discussion of these irregularities.}\\

\begin{figure}[h]
    \centering
    \includegraphics[scale=0.8]{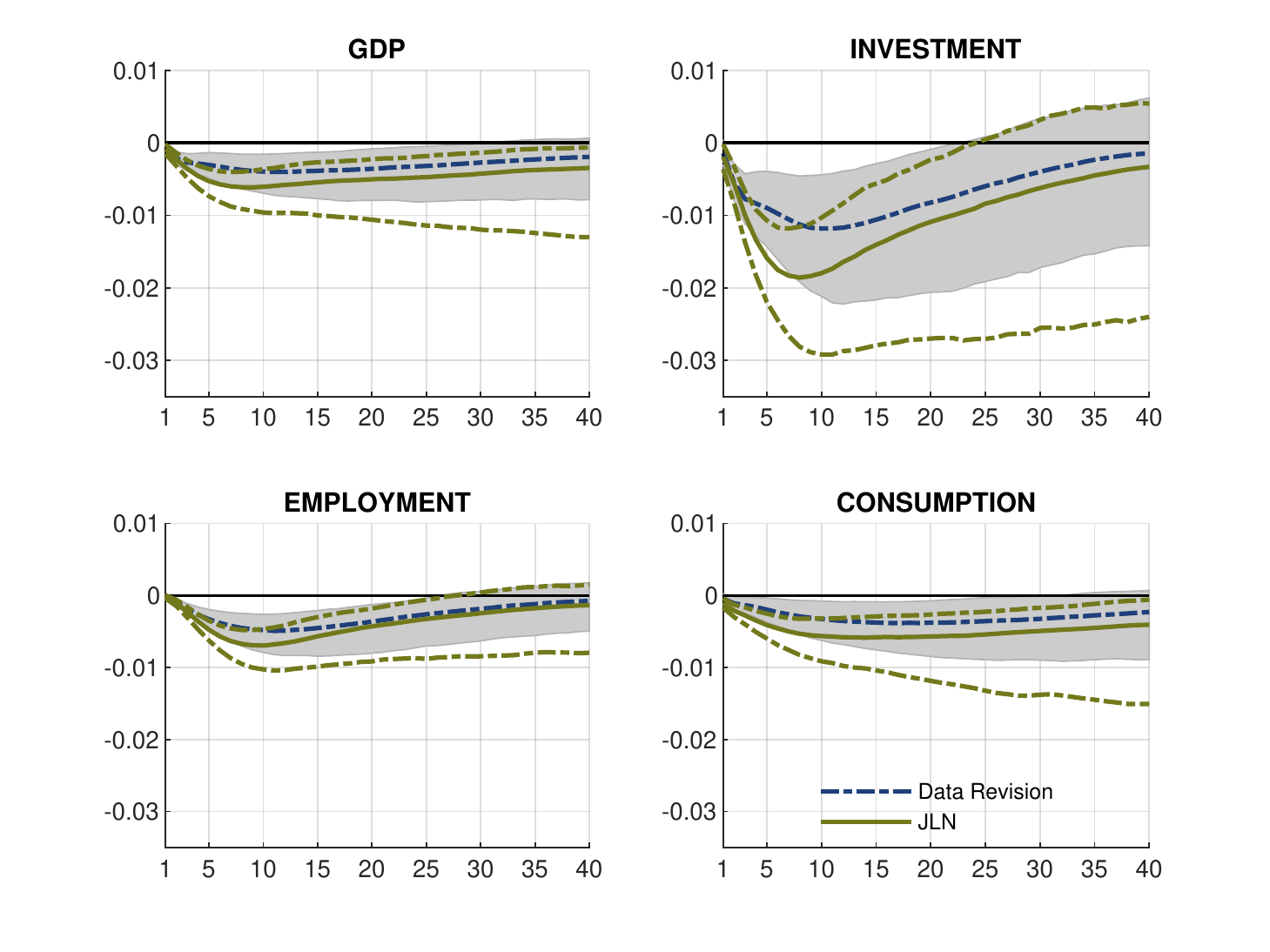}
    \caption{Impulse responses to an uncertainty shock in the U.S.
    \newline \scriptsize{\newline
			Notes: The dotted blue line depicts the posterior mean and the grey shaded area the 68\% error bands for the impulse responses to an one standard deviation uncertainty shock computed from a VAR model including the uncertainty measure based on data revisions. The solid green line depicts the posterior mean with the dotted green lines representing the 68\% error bands for the impulse responses from a model that uses the uncertainty measure of \cite{Juradoetal2015}. The estimation sample spans the period 1982Q1--2019Q4.} }
    \label{fig:irf_us}
\end{figure}

\noindent Figure \ref{fig:irf_us} reports the impulse responses of GDP, investment, employment and consumption to an uncertainty shock in the United States. A one standard deviation shock in uncertainty has an adverse and enduring effect on all macroeconomic variables (blue dashed line). For most of the variables the drop lasts for about two years, with most of posterior probability mass lying below zero. During this period output declines by around 0.4\%, investment by  1\% and, employment and consumption by about 0.5\%. The recovery from the uncertainty shock takes up to 10 years and more. While these effects seems somewhat strong and long-lasting, they are very similar to the uncertainty measure of \cite{Juradoetal2015} (green line). 

\subsection{Impact of Uncertainty Shock in G7 countries}\label{section: G7}
In this section, we examine the effects of uncertainty shocks within an international context. Thereby, we use a subset of the uncertainty indicators discussed in Section \ref{sec:macrouncertainty} to estimate the VAR model outlined above for the G7 countries. The intergovernmental economic organization comprises Canada, France, Germany, Italy, Japan, United Kingdom and the United States. In terms of economic importance, the organization makes up for about one third of global GDP based on purchasing power parity. Due to data limitations, we confine our estimates the sample from 1988Q1 to 2019Q4 for all countries. We employ the following country-specific VAR model
\begin{equation}
x_{i,t} = c_{i,b} + B_{i,1} x_{i,t-1} + B_{i,2} x_{i,t-2}  + ...+ B_{i,p} x_{i,t-p} + u_{i,t},  \label{eq: reducedform}
\end{equation}
where $x_{i,t}$ is a $n \times 1$ country-specific vector containing all $n$ endogenous
variables for country $i=1,...,N$ and time $t=1,...,T$. $c_{i,b}$ is an $n \times 1$ country specific fixed effect, $B_1,B_2...,B_p$ 
are $n \times n$ parameter matrices and $u_{i,t}$ is the $n \times 1$ disturbance with $u_{i,t}\sim N(0,\Sigma_{i})$, where $\Sigma_{i}$ is the $n \times n$ variance-covariance matrix. To obtain an aggregate impulse response for the G7 countries we average across country-specific impulse responses.\\

\begin{figure}[h]
    \centering
    \includegraphics[scale=0.8]{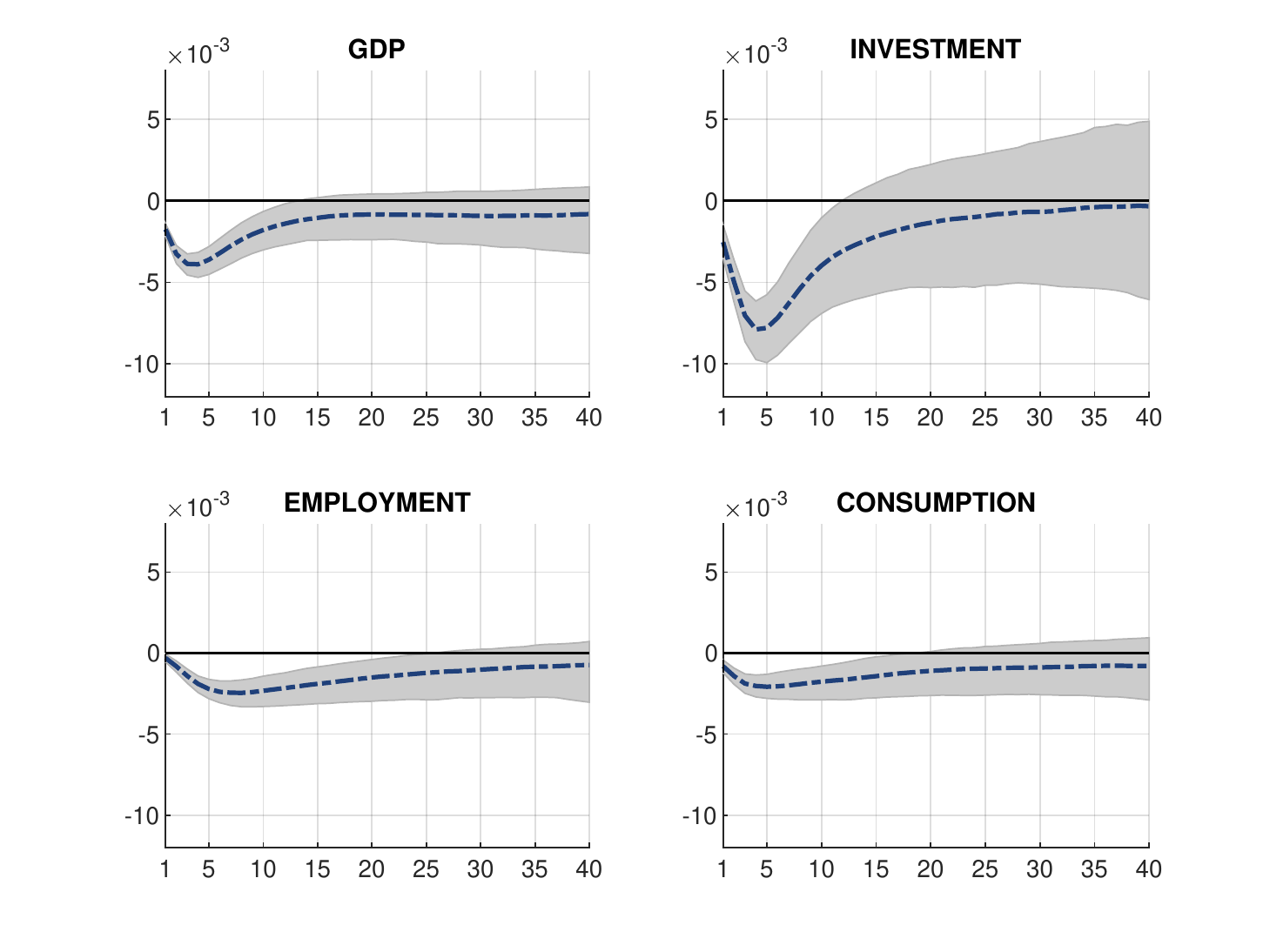}
    \caption{Impulse responses to an uncertainty shock for the group of G7 countries
    \newline \scriptsize{\newline
			Notes: The dotted blue line depicts the posterior mean and the grey shaded area the 68\% error bands for the impulse responses to an one standard deviation uncertainty shock. The estimation sample spans the period 1988Q1--2019Q4.} }
    \label{fig:irf_all}
\end{figure}

\noindent Figure \ref{fig:irf_all} shows the cross-country average impulse responses of GDP, investment, employment and consumption to an uncertainty shock. All variables unveil a negative relationship with uncertainty. While the impulse responses for the G7 aggregate in Figure \ref{fig:irf_all} are qualitatively similar to the impulse responses obtained for the United States reported in Figure \ref{fig:irf_us}, the average G7 effect of uncertainty shock is about half as strong as the one found for the United States.

\section{On the Role of Employment Protection Legislation}
Various studies have documented the negative impacts of uncertainty shocks on the labor market.\footnote{Studies documenting a negative effect of uncertainty on employment include, among others,  \cite{basu2017uncertainty}, \cite{Bloom2009}, \cite{bloom2018really}, \cite{caggiano2014uncertainty}, \cite{caggiano2017economic}, 
\cite{choi2015uncertainty}, \cite{caldara2016macroeconomic}, \cite{carriero2018measuring}, \cite{jo2019uncertainty}, \cite{Juradoetal2015}, \cite{leduc2016uncertainty}, \cite{mumtaz2018does}, \cite{netvsunajev2017uncertainty}, \cite{oh2019macro} and  \cite{scotti2016surprise}.} However, while the importance of the investment channel for the propagation of uncertainty shocks has been extensively studied in the literature, fewer studies focus on labor markets rigidities as the dominant transmission channel of uncertainty to the real economy. Recently, however, scholars started to explore this channel in more detail. \cite{cacciatore2015uncertainty}, for instance, show that binding downward rigidity of wages reinforce the negative effects of uncertainty on employment. In a similar fashion, \cite{leduc2016uncertainty} claim that nominal rigidities amplify the option-value channel through which uncertainty transmits the economy. \cite{guglielminetti2016labor} shows that firms reduce open vacancies when uncertainty increases in order to avoid expensive search activities and highlights its importance for the transmission of uncertainty shocks. \cite{matute2018uncertainty}, \cite{riegler2019impact} and \cite{jo2019uncertainty}  study the impact of uncertainty shock on labor flows. Summarizing, the authors find that uncertainty reduces hiring and increases lay offs and voluntary quits. In this study, we focus on the role of firing costs as a possible transmission mechanism of uncertainty shocks. We proxy firing costs with the degree of employment protection legislation and argue that stricter employment protection makes it more difficult---and thus more costly---to fire employees.\\

\noindent To obtain a better understanding of the role of employment protection legislation (EPL) in the propagation of uncertainty shocks, we first study the role of EPL within a theoretical framework and, in a second step, we use our newly developed uncertainty measures to empirically test the theoretical predictions. 

\subsection{Theoretical Model}\label{sec:theory}
To study the importance of EPL for uncertainty shocks within a theoretical model, we need a model that features uncertainty shocks and allows us to impose a stricter EPL. The dynamic stochastic general equilibrium model proposed by \cite{bloom2018really} includes these necessary features. The real business cycle model considers an economy with identical households wanting to maximize life-time discounted utility. All households choose how much they want to consume, work, and invest in order to maximize their life-time utilities. Furthermore, the model features an economy with heterogeneous firms that use labor and capital to produce a final good with the objective to maximize the life-time discounted value of their firm. Firms are subject to an exogenous process of productivity that has a firm-level and a macroeconomic component. Both the macroeconomic as well as the idiosyncratic productivity process vary in the first and second moment, with changes in second moment representing changes in uncertainty. Firms react to changes in productivity by adjusting capital and labor. However, adjusting capital and labor comes at a cost that firms have to take into account when maximizing their firm value.\\

\noindent We chose the model by \cite{bloom2018really} because of the exhaustive way to simultaneously model capital and labor adjustment costs. In our case, we are particularly interested in the way the authors model labor adjustment costs. The model includes two types of labor adjustment costs ($AC^{n}$). Firms face fixed and linear costs when adjusting labor. Fixed costs represent a lump sum cost that arises when employees are hired or fired. This cost does not depend on the size of the adjustment but on the state of the economy. One can think of these costs as arising from the deficiency in production owing to an experienced employee leaving the company or a new employee entering it. In contrast to fixed costs, linear costs depend on the size of the labor adjustment. These costs include, among others, recruiting and training costs for new employees and severance payment when laying off employees. Labor adjustment costs can thus be formally expressed as

\begin{equation}\label{eq:ac_n}
AC^{n} = \mathbbm{1}(|s|>0)y(z,A,k,n)C^{F}_{L}+\mathbbm{1}(s>0)C^{P}_{H}w+\mathbbm{1}(s<0)C^{P}_{F}w,
\end{equation}
	
\noindent where $C^{F}_{L}$ represent fixed labor adjustment costs that depend on the current state of production $y(z,A,k,n)$. $\mathbbm{1}(\cdot)$ represents an indicator function and $s$ indicates the change in employment. $C^{P}_{H}$ and $C^{P}_{F}$ represent hiring and firing costs as a percentage of the annual wage bill $w$.\\

\noindent As we aim to examine the effects of stricter employment protection legislation, we adjust the parameter that we associate most with stronger labor protection: firing costs. Stricter employment protection legislation makes it harder for firms to lay off employees. Hence, stricter employment protection legislation increases firing costs. Theoretically, firing costs change the effects of uncertainty on employment in two ways. First, an increase in firing costs reduces firing when uncertainty increases. Second, an increase in firing costs reduces hiring. The reason for this is the following: In the presence of non-convex adjustment costs---firing costs in our case---firms face Ss hiring/firing policy rules \citep{scarf1959optimality}. That is, firms do not hire new employees until productivity reaches an upper threshold (the S in Ss) and do not fire employees until its productivity hits a lower threshold (the s in Ss). Stricter employment protection legislation reduces the lower threshold. Hence, productivity needs to fall more before firms start firing employees. Overall employment will fall less compared to an economy with lower employment protection standards. This mechanism is similar to the one described by \cite{bell1996adjustment}. Using a partial equilibrium model, the author shows that an increase in firing costs has a negative effect on employment because firms reduce hiring due to precautionary reasons. Overall, however, the negative effect on uncertainty is reversed as the increase in costs discourages firing by more than it does hiring. The reason being that laying off employees causes immediate costs, while hiring costs are discounted as they only become relevant once a firm fills the vacant position again. \\

\noindent In order to study the role of EPL the propagation of uncertainty shocks we calibrate, solve and simulate the model of \cite{bloom2018really} twice, once for an economy for low EPL and once for an economy with high EPL. In order to ensure comparability with \cite{bloom2018really} and the RBC literature in general, we do not change the calibration proposed by \cite{bloom2018really}, except for the firing cost parameter. For the United States, \cite{bloom2018really} assume firing costs to be on average 1.8\% of an annual wage bill. For continental European economies---countries that according to the OECD Employment Protection Database have on average stricter EPL---studies report substantially higher firing costs \citep{grund2006severance,kramarz2010shape}. It is surprisingly hard to find studies quantifying firing costs for countries other than the US. \cite{del1998much} interview Italian manufacturing firms and find firing costs that range from less than half a monthly (3.6\% of an annual wage bill) of labour cost to up to 20 months of labour costs (166\% of an annual wage bill) in cases of a conflict. Unfortunately, the authors do not provide averages. For illustrative purposes, we start from the lowest reported value that is twice as high as in the case of \cite{bloom2018really}, i.e. we assume firing costs to be on average 3.6\% of an annual wage bill. Table \ref{t:dsge} summarizes the labor adjustment parameters.\\
	
\begin{table}[h]
	\centering
	\scalebox{0.90}{
		\begin{tabular}{lp{80mm}cp{40mm}}
			\multicolumn{2}{l}{Parameter Description}				& Low EPL & High EPL    \\ \hline
			&		       											& 				& 	\\ 
			$C^{F}_{L}$:	& fixed hiring/firing costs (\% sales)  & 0.021 		& 0.021 \\
			$C^{P}_{H}$: 	& per capita hiring (\% of annual wage bill)		    & 0.018			& 0.018\\
			$C^{P}_{F}$: 	& per capita firing cost (\% of annual wage bill)		& 0.018			& 0.036\\
		\end{tabular}
	}
	\caption{Model Calibration \newline \scriptsize{\newline
			Notes: This table presents the model calibration and parameter choices. The calibration reflects a quarterly calibration of the model and is based on \cite{bloom2018really}.}
	}\label{t:dsge}
\end{table}

\begin{figure}[h]
    \centering
    \scalebox{1}{
    \includegraphics{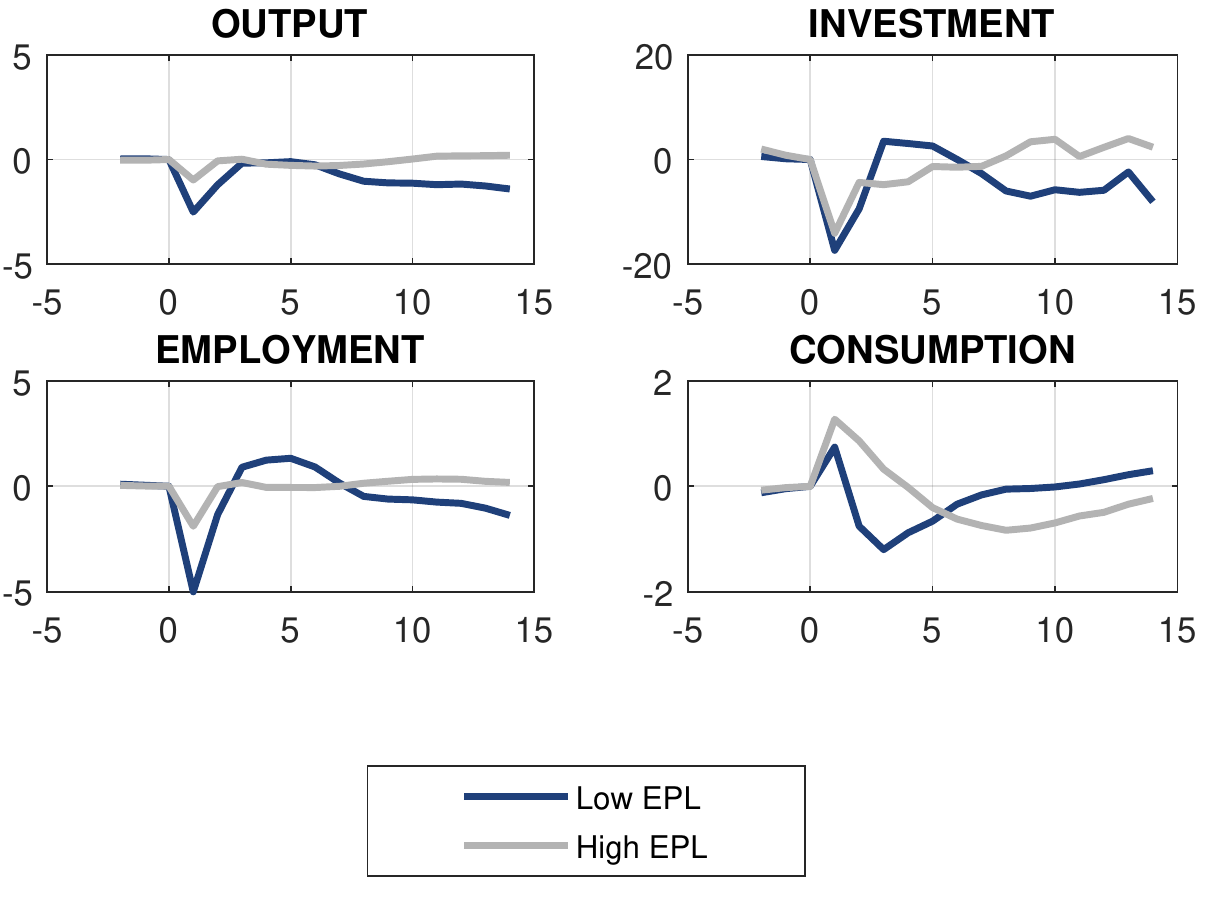}
    }
    \vspace{-0cm}\caption{Uncertainty Shocks under high and low labor protection \newline \scriptsize{\newline
						Notes: This figure presents DSGE impulse responses of output, investment, employment and consumption after an uncertainty shock. Thereby, the blue lines presents the impulse response to uncertainty shock under low employment protection legislation and the gray line shows the impulse response of the same uncertainty shock under high employment protection.}
		}\label{fig:irfDSGE}
\end{figure}

\noindent We use two differently calibrated models to simulate the reaction of the economy to an uncertainty shock. In the case of this model, an uncertainty shock corresponds to an increase in variance of the shock distribution from which future realisations of productivity (TFP) will be drawn. Figure \ref{fig:irfDSGE} presents the impulse responses of output, investment, employment and consumption to an uncertainty shock. The blue lines presents the impulse response of an uncertainty shock under low employment protection legislation and the gray line shows the impulse response of the same uncertainty shock under high employment protection. Both economies contract after an uncertainty shock. The uncertainty shock causes an immediate contraction of output in the first period followed by a recovery starting in the subsequent quarters. Three channels contribute to this fall in output: the investment channel, the employment channel, and the misallocation of factors of production.\footnote{Section 5 of \cite{bloom2018really} provides an in-depth discussion of these channels and dissects the impact of an uncertainty shock in its various components.} The presence of capital adjustment costs causes investment to drop sharply after an uncertainty shock. In the first period investment drops by around 15\% followed by a rapid recovery. Also employment reacts negatively to an uncertainty shocks. As firms face firing and hiring costs when adjusting their number of employees, an increase in uncertainty reduces hiring and firing activities of firms. Thereby, hiring decreases more than firing. Moreover, employment also decreases because of labor attrition. Finally, the decrease in firing and hiring increases the misallocation of factors of production. Households expect the increase in misallocation to decrease future productivity and thus to lower the expected return on savings making immediate consumption more attractive. These dynamics lead in combination with the available resources in the economy---the capital stock does not adjust to uncertainty in period one---to an increase in consumption in the first period. The increase of consumption following an uncertainty is a well-known artefact of this model. \cite{bloom2018really} extensively discuss this behaviour in their paper. As we focus on role of firing cost on the propagation of uncertainty shocks, we are primarily interested in changes of the consumption dynamics relative to the original specification.\\

\noindent When increasing firing costs the reaction of output, investment, employment, and consumption to an uncertainty shock remain similar in their dynamics. However, the amplitude of the time profile changes considerably. Increasing firing costs increases the option value of waiting as it makes it more expensive for firms to lay off employees. Thus, the uncertainty shocks reduces employment less compared to the original specification. In contrast to employment, investment still decreases almost by as much after the uncertainty shocks as in the original specification. As more production factors remain in the economy, output reduces by less than in the original specification. Finally, the increase in firing costs increases the misallocation of factor of production which in combination with higher production increases consumption compared to the original specification. Overall, increasing firing costs causes an uncertainty shock to have less contractionary effects on the economy.

\subsection{Evidence from a VAR}
\noindent Now, we use the international set of revision based measures of uncertainty to test this theoretical prediction. We thus split countries into two groups according to their strictness of employment protection.\footnote{We confine this analysis to countries for which we have data since 1990. Hence, we end up with the United States, Canada, United Kingdom, Switzerland, Japan, France, Germany, Sweden and Italy.} To split countries according to their degree of employment protection, we use the annual time series data of the OECD Employment Protection Database to calculate the average value of the strictness of employment protection. Specifically, we use the measure of individual and collective dismissals (EPRC\_V1) from 1985 to 2013. Table \ref{t:epl} ranks countries according to the strictness of employment protection. \\

% \begin{table}[!htbp]
% 	\centering
% 	\scalebox{0.90}{
% 		\begin{tabular}{lc} 
% 			Country         &	Average EPL (1985-2013)   \\ \hline  
% 				            &     \vspace{-0.5cm}       \\ 
% 			United States   &    0.26    \\ 
% 			Canada          &    0.92    \\ 
% 			United Kingdom  &    1.17    \\ 
% 			Australia       &    1.32    \\ 
% 		    Switzerland     &    1.60    \vspace{-0.5cm}\\ 
% 			                &            \\ 
% 		    \textbf{Median}          &    1.61    \vspace{-0.5cm}\\ 
% 			                &            \\ 
% 			Japan           &    1.62    \\ 
% 			France          &    2.39    \\ 
% 			Germany         &    2.65    \\ 
% 			Sweden          &    2.70    \\ 
% 			Italy           &    2.76    \\ 

% 		\end{tabular}
% 	}
% 	\caption{High EPL vs. Low EPL countries \newline \scriptsize{\newline
% 			Notes: This table ranks countries according to the strictness of employment protection. In order to calculate the ranking, we use the annual time series data of the OECD Employment Protection Database to calculate the average value of the strictness of employment protection – individual and collective dismissals (EPRC\_V1) - over time (from 1985 to 2013).}
% 	}\label{t:epl}
% \end{table}

\begin{table}[h]
	\centering
	\scalebox{0.90}{
		\begin{tabular}{lc} 
			Country         &	Average EPL (1985-2013)   \\ \hline  
				            &     \vspace{-0.5cm}       \\ 
			United States   &    0.26    \\ 
			Canada          &    0.92    \\ 
			United Kingdom  &    1.17    \\ 
		    Switzerland     &    1.60    \vspace{-0.5cm}\\ 
			                &            \\ 
		    \textbf{Median}          &    1.62    \vspace{-0.5cm}\\ 
			                &            \\ 
			Japan           &    1.62    \\ 
			France          &    2.39    \\ 
			Germany         &    2.65    \\ 
			Sweden          &    2.70    \\ 
			Italy           &    2.76    \\ 

		\end{tabular}
	}
	\caption{High EPL vs. low EPL countries \newline \scriptsize{\newline
			Notes: This table ranks countries according to the strictness of employment protection. In order to calculate the ranking, we use the annual time series data of the OECD Employment Protection Database to calculate the average value of the strictness of employment protection – individual and collective dismissals (EPRC\_V1) - over time (from 1985 to 2013).}
	}\label{t:epl}
\end{table}

\begin{figure}[h]
    \centering
    \includegraphics[scale=1]{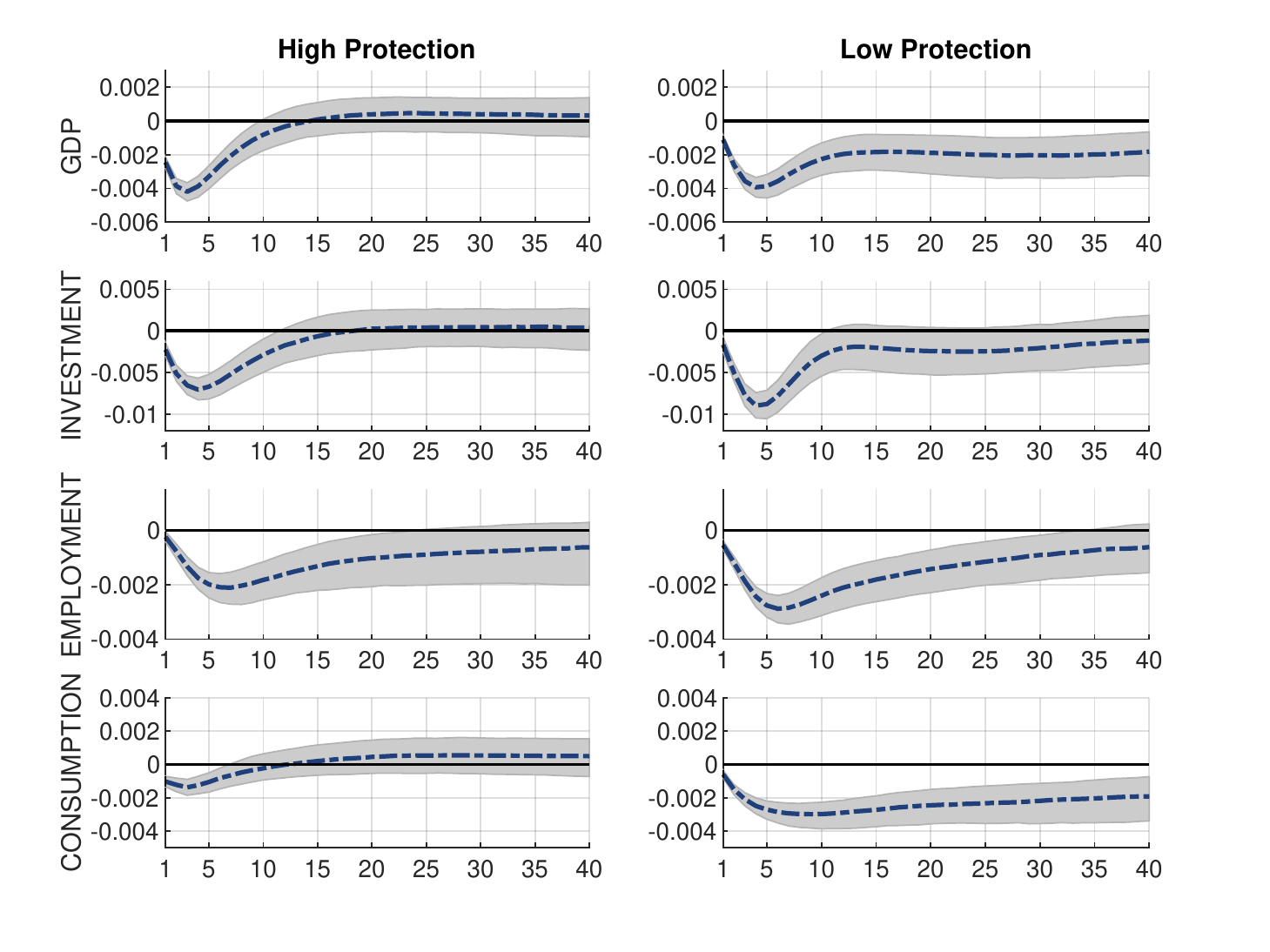}
        \caption{Impulse responses to an uncertainty shock for high EPL countries (left panel) and low EPL countries (right panel).
    \newline \scriptsize{\newline
			Notes: The dotted blue line depicts the posterior mean and the grey shaded area the 68\% error bands for the impulse responses to an one standard deviation uncertainty shock. The estimation sample spans the period 1988Q1--2019Q4.} }
    \label{fig:irf_hl}
\end{figure}

\noindent The groups selected in Table \ref{t:epl} mirror our expectations with Anglo-Saxon economies displaying a low degree and continental European countries showing higher degree of employment protection standards. According to OECD Employment Protection Database, Switzerland and Japan have a very similar degree of EPL. In our baseline specification, we include Switzerland in the group with low labor protection and Japan in the group of high EPL countries.\footnote{As a robustness test, we re-run the analysis excluding both Japan and Switzerland. Neglecting the two countries does not significantly change the results (see Figure \ref{fig:irf_hl_short} in Appendix \ref{sec:app__rob}).} To obtain group-specific impulse responses to an uncertainty shock, we use the VAR described in section \ref{section: G7}, averaging across country-specific impulse responses for each group. Our empirical findings suggest that uncertainty has indeed less deteriorating effects in countries with high employment protection. Figure \ref{fig:irf_hl} shows that the effect of a one standard deviation uncertainty shock is not only more contractionary in countries with high protection compared to countries with low labor protection, the negative effects are also more persistent.\\

\noindent The results presented in Figure \ref{fig:irf_hl} are consistent with the theoretical predictions outlined above. In countries with stricter employment protection legislation, it is more costly for firms to reduce employment. Hence, employment drops less in the light of an uncertainty shock.\footnote{We divide countries according to their level of employment protection legislation. Theoretically, the two groups of countries could also differ along other dimension that are relevant for the uncertainty transmission channel, i.e. level of capital adjustment costs. However, we cannot think of a good reason why a country's employment protection legislation should correlate with its capital adjustment costs. In the absence of comparable capital adjustment cost data, we leave it to future research to control for capital adjustment costs.} Because of the lower response of employment and because of the complementarity of capital and labor, firms do not cut investment by as much, causing production to contract less. Finally, uncertainty decreases consumption less in countries with high labor protection. As pointed out in \ref{sec:theory}, an uncertainty shock increases the misallocation in an economy more if firing costs are higher. Households expect the increased missalocation to decrease future return on savings and they decrease consumption less. While this mechanism might explain parts of the differences in the consumption dynamics, precautionary saving might also be part of the story. A higher degree of labor protection transmits into a higher job security of employees. An increase in uncertainty lets employees worry less about their future income in case of high employment protection than in case of low employment protection. Households thus increase precautionary saving less and decrease consumption by less which causes a less pronounced drop of aggregate demand. Our findings are consistent with the evidence presented by \cite{jo2019uncertainty} that indicates that compared to Germany, labor market frictions in the U.S. might be too low for uncertainty to have strong real option effects on employment. Our evidence complements the literature on the importance of the labor channel in explaining the transmission of uncertainty shocks. However, it highlights a different mechanism. In contrast to \cite{guglielminetti2016labor}, who argues for the importance of hiring costs, our results indicate a prominent role of firing costs in explaining the dynamics of uncertainty shocks.\\ %From a theoretical point of view, firing and hiring costs coexist independently next to each other. The combination of both channels determines the importance of the labor channel in the propagation of uncertainty shocks in the model. In a situation where firing costs are very high and hiring are zero, the labor channel would be almost irrelevant. On the other hand, in an economy with no firing costs and high hiring costs the labor channel is an important transmission mechanism.\\ 

\section{Conclusion}
In this paper we have introduced new internationally comparable measures of macroeconomic uncertainty for a large set of countries using data revisions in aggregate variable that are bound to the system of national accounts. We have set up an econometric model and constructed a new real-time data set of real GDP for 39 countries that serves as the basis for our estimations. Using real-time data permits us to obtain accurate estimates of forecast error based uncertainty that an economic agent experienced at any given point in time, whereas existing measures of macroeconomic uncertainty base on forecast errors that are constructed with non real-time data.\\%These non real-time data estimates thus include observations that exceeds the economic agents information sets at any given point in time.\\ 

%Compared to existing measures of macroeconomic uncertainty, this approach provides estimates of macroeconomic uncertainty that are closer to the uncertainty that economic agents actually perceived at any given point in time.\\

\noindent In order to obtain real-time uncertainty estimates, we extended the data revision model proposed by \cite{JacobsvanNorden2011} such that it allows us to extract the volatility of the unpredictable part of future releases of the news component that forms our measure of macroeconomic uncertainty. We showed that the resulting uncertainty indicator for the United States has similar properties than the macroeconomic uncertainty measures proposed by \cite{Juradoetal2015}. The revision based indicator is thereby less volatile than alternative measures such as the economic policy uncertainty index by \cite{baker2016measuring} or the VIX and the revision based indicator also identifies the same three major uncertainty shocks between 1965 and 2016. Namely, the recession in the 1970s, the early 1980s recession and the Great Recession of 2008. The revision based indicator reaches its highest peak during 1970s. Considering that the recession in the 1970s comprised the first oil price shocks, the collapse of the Bretton Woods System and the end of the post World War II economic expansion this seems coherent with a broader economic history perspective. Our empirical evaluation indicates a strong and negative relationship between the revision based uncertainty measures and the economy. Estimating VARs for the United States and the G7 countries shows that a one standard deviation shock in the revision based uncertainty indicators leads to a contraction in GDP, investment, employment and consumption.\\

\noindent We studied the importance of labor market frictions for the propagation of uncertainty shocks. In a cross-country VAR analysis, we found that uncertainty shocks have more deteriorating effects in countries with a lower degree of EPL compared to countries with stricter EPL. Using the theoretical model of \cite{bloom2018really} with varying degree of firing costs, we show that these empirical findings are in line with theory.

\clearpage
\bibliography{literature}
\begin{appendices}
\section{Posterior Simulations}\label{sec:PosteriorSimulation}
In this section we describe the blocks of our Gibbs sampling procedure outlined in section \ref{sec:econometric}. Note that Equations (\ref{eq:measure}) and (\ref{eq:state3}) can be rewritten as
\begin{align}
Y_{t}&=%
Z%
\alpha_{t} \label{eq:measure4},\\
\alpha_{t} &=%
\varphi_{t} +%
T_{t} %
\alpha_{t-1}  +%
R D_{t}^{1/2}%
\eta_{t}, \label{eq:state4}
\end{align}%%
with 
\begin{align}
R=%
\begin{bmatrix}
1 & 1 & 0 & 0 \\ 0 & -1 & 0 & 0  \\ 0 & 0 & 0 & 0 \\ 0 & 0 &  1 & 0 \\ 0 & 0 & 0 & 1
\end{bmatrix}, \nonumber
D_t=%
\begin{bmatrix}
\exp(h_{t}^{\nu 1 }) & 0 & 0 & 0 \\ 0 & \exp(h_{t}^{\nu L}) & 0 & 0 \\ 0 & 0 &  \exp(h_{t}^{\zeta 1}) & 0 \\ 0 & 0 & 0 & \exp(h_{t}^{\zeta L})
\end{bmatrix}. \nonumber 
\end{align}

% \ \\
% \subsection{Estimation Routine}
% We sample the posterior distribution of our model by repeating the following routine
% \begin{enumerate}
%     \item Use Kalman filter to recover $\alpha_{t-1}$ conditional on $\varphi_{t}$, $T_{t}$ and $D_{t}$ and calculate the residuals $\nu_{t}$
%     \item Update the stochastic volatilities $D_{t}$ conditional on the residuals $\nu_{t}$
%     \item Jointly update $\varphi_{t}$ and $T_{t}$ conditional on $D_{t}$, $\alpha_{t-1}$
% \end{enumerate}

\subsection{Drawing News, Noise and True Values}
Conditional on the stochastic volatilities and the VAR coefficients of the state space model ($c_t$ and $\rho_t$), we draw $\alpha_t$ for $t=1,..,T$, using the forward filtering backward sampling procedure of \cite{CarterKohn1994} and \cite{Fruhwirth-Schnatter1994}, where (\ref{eq:measure4}) serves as observation equation and (\ref{eq:state4}) as state equation.%From that we recover the residual $u^{x}_{t}$ with $x=\tilde{y},\nu,\zeta 1,\zeta L$. \\

\subsection{Drawing Stochastic Volatility}
\noindent Conditional on news, noise, true values and the VAR coefficients of (\ref{eq:state4}), we obtain draws for the stochastic volatilities ($h_{t}^{x}$) using the estimation method proposed in \cite{kastner2014ancillarity}. We thus estimate the stochastic volatilities by interweaving estimation models specified in a centered (C) and a non-centered (NC) parameterization using the ancillarity-sufficiency interweaving strategy (ASIS) detailed by \cite{yu2011center}. This estimation strategy addresses the trade-off in terms of sampling efficiency depending on the value of the persistence parameter $\phi$. \cite{kastner2014ancillarity} show that this estimation strategy outperforms estimators using pure centered or non-centered parameterizations.\\

\noindent Starting from Equation (\ref{eq:state2}), we assume that:

\begin{align}
\sigma^{i}_{t}\eta^{i}_{t}  &= e^{h_{t}^{i}/2}\eta^{i}_{t}
\label{eq:measure_vola}
\\
h_{t}^{i}  &= \mu^i + \phi^i (h_{t-1}^{i} - \mu^i) + \tau^{i}\epsilon_{t}^{i} \text{\ \ },
\label{eq:state_vola2}
\end{align}

\noindent with $\epsilon_{t}^{i} \sim N(0,1)$ and $x=\nu 1,\nu L,\zeta 1,\zeta L$. One can express the centered parametrization of Equation (\ref{eq:state_vola2}) as a non-centered parametrization. In the non-centered parametrization the volatility components are assumed to follow the following process:

\begin{equation}
\tilde{h}_{t}^{i}  = \phi^{i} \tilde{h}_{t-1}^{i} + \epsilon_{t}^{i},
\end{equation}
with $\epsilon_{t}^{i} \sim N(0,1)$ and $x=\nu 1,\nu L,\zeta 1,\zeta L$.\\

\noindent We can rewrite Equation (\ref{eq:measure_vola}) as
\begin{equation}\label{eq:state_vola2}
    \tilde{\sigma}^{i}_{t} = h^{i}_{t} + log((\eta^{i}_{t})^{2})
\end{equation}
with $\eta^{i}_{t} \sim N(0,1)$ and $\tilde{\sigma}^{i}_{t}$ denotes $log((\sigma^{i}_{t}\eta^{i}_{t})^{2})$. The fact that we can approximate the distribution of $log((\eta^{i}_{t})^{2})$ by a mixture of normal distributions, that is, $log((\eta^{i}_{t})^{2})|r^{i}_{t} \sim N(m^{i}_{r_{t}},(s^{i})^{2}_{r_{t}})$ with $r_{t}$ indicating the mixture component, we can rewrite Equation (\ref{eq:state_vola2}) as a linear and conditionally Gaussian state space model,

\begin{equation}\label{eq:state_vola_linear}
    \tilde{\sigma}^{i}_{t} = m_{r_{t}}^{i}+h^{i}_{t} + \eta^{i}_{t},
\end{equation}

\noindent with $\eta^{i}_{t} \sim N(0,s^{{2}^{i}_{r_{t}}})$. Based on Equation (\ref{eq:state_vola_linear}), we apply a MCMC procedure outlined in \cite{kastner2014ancillarity} that interweaves the centered and non-centered specification. %In order to take into account the relative efficiency of the centered and non-centered specification, the authors propose an interweaving strategy. When estimating stochastic volatilities one faces a trade-off in terms of sampling efficiency depending on the value of the persistence parameter $\phi$. That is, conditional on the value of $\phi$ either the centered specification or the non-centered specification becomes more efficient. In case $\phi$ is close to zero, the latent states (h) become very informative about $\mu$ once the the conditional variance $\eta^{2}$ becomes small. Thus, in case $\phi$ is close to zero and $\sigma^2$ is small, the centered specification becomes inefficient with respect to the non-centered specification. The reason being that treating h as latent data implies that more information is missing. On the contrary, we expect the centered specification to become more efficient once the persistence parameter $\phi$ get close to 1. Once the latent process approaches a random walk, h will become relatively uninformative about $\mu$ and comparatively little information is lost when treating h as latent data.\\

%\noindent Unfortunately, the likelihood function of stochastic volatility models with discretely sampled data has usually an intractable form. In applying Bayesian estimation methods, one can overcome this problem by sampling the latent states (h) and treat them as known for updating the parameters $\mu$, $\phi$ and $\eta$. This demands the specification of a prior distribution $p(\mu,\phi,\sigma)$. Thereby, 
\noindent We rely on the priors proposed in \cite{kastner2014ancillarity}. That is, $\mu$ follows a normal distribution with mean $b_{\mu}$ and variance $B_{\mu}$, i.e. $\mu \sim N(b_{\mu},B_{\mu})$. The persistence parameter $\phi$ follows a Beta distribution, i.e. $B(a_{0},b_{0})$. Finally, for the volatility parameter $\sigma$, they chose $\sigma^{2} \sim B_{\sigma}\chi^{2}_{1}=G(1/2,1/2B_{\sigma})$. We use the same priors for the centered and non-centered parameterization. We calibrate the parameters as follows: $b_{\mu}=0$, $B_{\mu}=100$, $a_{0}=5$, $b_{0}=1.5$ and $B_{\sigma}=1$.\\

\noindent The MCMC interweaving procedure consists of the following steps:

\begin{enumerate}
   \item We sample the volatilities using the all without a loop (AWOL) procedure by drawing from $h_{[-0]}^{i}|\tilde{\sigma}^{i},r^{i},\mu^{i},\phi^{i},(\eta^{i})^{2}$. The initial values are drawn from $h_{0}^{i}|h_{1}^{i},r^{i},\mu^{i},\phi^{i},(\eta^{i})^{2}$.
    \item We sample $\mu^{i},\phi^{i}$, and $(\eta^{i})^{2}$ using Bayesian regression. We use a 2-block sampler, where we draw $(\eta^{i})^{2}$ from $(\eta^{i})^{2}|h^{i},\mu^{i},\phi^{i}$ and we jointly sample $\mu^{i}$ and $\phi^{i}$ from $\mu^{i},\phi^{i}|h^{i},(\eta^{i})^{2}$. Because the chosen priors are not analytically tractable, we calculate updates via a Metropolis-Hastings (MH) algorithm.
    \item Move to NC by the deterministic transformation $\tilde{h_{t}^{i}}=\frac{h_{t}^{i}-\mu^{i}}{\eta^{i}}$
    \item Redraw $\mu^{i}, \phi^{i}, (\eta^{i})^{2}$ for NC specification. We need MH only to update $\phi^{i}$ by drawing from $\phi^{i}|\tilde{h}^{i}$. We can Gibbs-update $\mu^{i}$ and $(\eta^{i})^{2}$ jointly from $\mu^{i}, (\eta^{i})^{2}|\tilde{\sigma}^{i},\tilde{h}^{i},r^{i}$.
    \item Move back to C by calculating $h_{t}^{i}=\mu^{i}+\eta^{i} \tilde{h}_{t}^{i}$ for all t
    \item Draw the indicators $r^{i}$ (C). We update the mixture component indicators $r^{i}$ from $r^{i}|\tilde{\sigma}^{i}$ using inverse transform sampling.
\end{enumerate}
\subsection{Drawing Time-Varying Coefficients}
Conditional on news, noise, true values and the stochastic volatilities, we draw the time-varying coefficients of Equation (\ref{eq:state4}), by considering the process of true GDP $\tilde{y}_{t}$ to be represented by the following state space model

\begin{align}
\tilde{y}_{t}&=%
Z_{t}%
\alpha_{t} + G_{t}u_{t}\label{eq:measure2},\\
\alpha_{t} &=%
T %
\alpha_{t-1}  +%
H_{i,j} e_{t}%
\end{align}%%

with

\begin{align}
Z_{t}= 
\begin{bmatrix}
1 & \tilde{y}_{t}
 \end{bmatrix},
\alpha_{t} = 
\begin{bmatrix}
c_{t} \\ \rho_{t}
 \end{bmatrix}
G_{t}=  %
\begin{bmatrix}
 \exp(h_{t}^{\nu 1 }) & \exp(h_{t}^{\nu L})
\end{bmatrix},\\
u_{t}=  %
\begin{bmatrix}
 u_{t}^{\nu 1 } \\ u_{t}^{\nu L}
\end{bmatrix}, \nonumber 
T =  %
\begin{bmatrix}
 1 & 0\\ 
 0 & 1 
\end{bmatrix},
H_{i,j}=  %
\begin{bmatrix}
 \sigma_{i,j}^{c_{i,j}} & \sigma_{i,j}^{\rho_{i,j}}
\end{bmatrix},
e_{t}=  %
\begin{bmatrix}
 e_{t}^{c_{t}} \\ e_{t}^{\rho_{t}}
\end{bmatrix}. \nonumber
\end{align}%%

\noindent We obtain estimates for $c_{t}$ and $\rho_{t}$ using the method proposed by \cite{McCauslandMillerPelletier2011}. We set up the diagonal matrix $\Omega$ as follows:

\textbf{\underline{Diagonal Matrix $\Omega$}}
\begin{align}
\Omega=%
\begin{bmatrix}
\Omega_{11} & \Omega_{t/t-1} & 0 & \cdots & 0 \\ 
\Omega_{t/t-1} & \Omega_{tt} & \Omega_{t/t-1} & \ddots & \vdots \\ 
0 & \Omega_{t/t-1} &  \ddots & \ddots & 0 \\ 
\vdots & \ddots & \ddots &  \Omega_{tt} & \Omega_{t/t-1}  \\
0 & \cdots & 0 & \Omega_{t/t-1} & \Omega_{nn} \\
\end{bmatrix}
\end{align}

\textbf{\underline{Diagonal Elements}}
\begin{align}
\Omega_{11}&= 
Z_{1}'A_{11,1}Z_{1}+A_{22,1}+P_{1}^{-1}\\
\Omega_{tt}&= 
Z_{t}'A_{11,t}Z_{t}+2A_{22,t}\\
\Omega_{nn}&= 
Z_{n}'A_{11,n}Z_{n}+A_{22,n}\\
\end{align}%%

\textbf{\underline{Off-Diagonal Elements}}
\begin{align}
\Omega_{t/t-1}&= 
-A_{22,t}\\
\end{align}%%

with 

\begin{align}
A_{t}&= 
\begin{bmatrix}
(G_{t} \ '\  G_{t})^{-1} & 0 \\
0 & (H_{i,j} \ '\  H_{i,j})^{-1} 
 \end{bmatrix}\\
\end{align}%%

We set up the co-vector $C$ the following way:

\textbf{\underline{Covector}}
\begin{align}
C_{1}=%
\begin{bmatrix}
C_{1}\\ 
C_{tt}\\ 
\vdots\\ 
C_{nn}\\ 
\end{bmatrix}.
\end{align}

with
\begin{align}
C_{1}&= 
Z_{0}'A_{11,0}\tilde{y}_{0}+P_{1}^{-1}a_{1}\\
C_{tt}&= 
Z_{t}'A_{11,t}\tilde{y}_{t}\\
C_{nn}&= 
Z_{T}'A_{11,T}\tilde{y}_{T}
\end{align}%%

\noindent We solve this system by first computing the Cholesky decomposition $\Omega=LL'$ that we implement directly in Julia.  Because of the band structure of $\Omega$, the decomposition is incredibly fast. In a second step, we draw $e_{t} \sim N(0,1)$ and solve $La =c$ for $a$. Finally, we use back-substitution in order to solve $L'h=a+e$ for $h$.

\section{Real-Time Data}\label{sec:app__real_time_data}
We use data revisions in macroeconomic aggregates to obtain measures of macroeconomic uncertainty for various OECD countries. Real-time data releases of macroeconomic aggregates are thereby the key ingredient to construct the uncertainty indicator. In our preferred specification, we base the indicator on nominal GDP. In order to obtain a comprehensive data set for various countries, we need to tap and combine several data sources. Appendix \ref{sec:app__real_time_data} describes these various data sources and outlines the construction of our data base in great detail.\\

\subsection{Real-Time Data: Main Source}

\noindent The largest part of our data is provided by \textit{Original Release Data and Revisions Database}. The \textit{Original Release Data and Revisions Database} is part of OECD Main Economic Indicators database \citep{oecd2017mei} and represents the central data source of this project. The database provides different releases of macroeconomic aggregates for many countries. This study uses data from 32 countries including Australia, Austria, Belgium, Brazil, Canada, Denmark, Finland, France, Germany, Great Britain, Greece, Iceland, India, Indonesia, Ireland, Israel, Italy, Japan, Korea, Luxembourg, Mexico, Netherlands, New Zealand, Norway, Portugal, Russia, South Africa, Spain, Sweden, Switzerland, Turkey, USA. Unfortunately, \textit{Original Release Data and Revisions Database} provides releases of macroeconomic variables only since 1999. For data prior to 1999, we need to rely on other data providers. We primarily use the data made available by the Federal Reserve of Dallas for releases prior to 1999. \cite{fernandez2011real} collect real-time data for various economies including those that we use in this study. The authors assemble the dataset from original quarterly releases of different macroeconomic aggregates from 1962 to 1998. We use their data for all countries except the Australia, New Zealand, and U.S.. For the U.S., we use data provided by the Federal Reserve of Philadelphia as they provide more exhaustive data compared to the data provided by Federal Reserve of Dallas. We use Australian nominal GDP data provided by the Australian Real-Time Macroeconomic Database (\cite{lee2012australian}). For New Zealand, we use the ``Real-Time GDP Data'' data set from the Reserve Bank of New Zealand that provides nominal GDP releases. Table \ref{tab:country_overview} provides an overview of the countries included in our study, the data provider and the first available data point. \\

\begin{figure}[!htbp]
    \centering
    \includegraphics[scale=0.8]{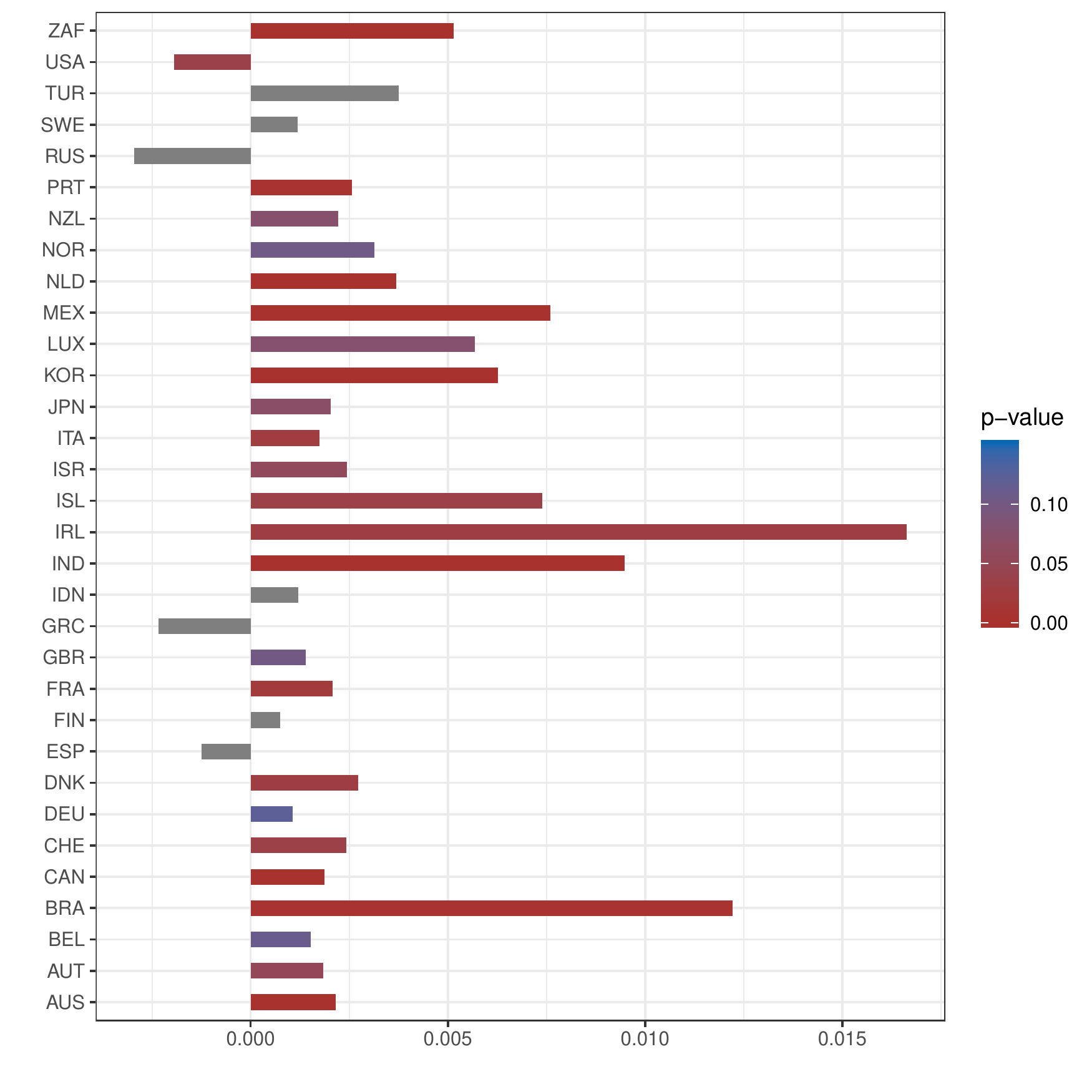}
    \caption{Real GDP Growth (yoy): Mean of the 10th revisions for the period from 2000 Q1 to 2016Q3.}
    \label{fig:diverge_bars}
\end{figure}

\noindent Figure \ref{fig:diverge_bars} shows the average of the 10th revision of year-over-year growth rates of nominal GDP. Thereby, most countries have a statistical significant downward bias. That is, ten quarter after the first release, most countries countries publish on average a higher growth rate. 

\subsection{Real-Time Data: Practical decisions}

\noindent\underline{Australia}\\
\noindent We use real-time data from the Real-Time Macroeconomic Database provided by the University Melbourne Macroeconomics Research Unit. The 1973Q3 vintage contains unusual  entries.  The  1973Q2  vintage  contains  combined  figures for the 1972Q3-Q4 and 1973Q1-Q2 reference dates. We split the combined figures by using the share of the 1973Q4 vintage.

\noindent\underline{Canada}\\
\noindent We use real-time data provided by the Federal Reserve of Dallas for releases prior to 1982. For the time span between 1982 and 1999, we rely on real-time data provided by the Bank of Canada. Finally, from 1999 onward, we use real-time data provided by the OECD Main Economic Indicators database. The data provided by the OECD Main Economic Indicators database and the Bank of England are virtually identical after 1999, expect for the quarters 2000Q2 to 2001Q1. For this quarter, we opt for data provided by the Bank of Canada as they appear to be by far less volatile. \\

\noindent\underline{Germany} \\
\noindent We use real-time data provided by the Federal Reserve of Dallas for releases prior to 1999. The dataset provides GNP until February 1993. Thereafter the dataset provided by the Federal Reserve of Dallas contains data on GDP. Furthermore, the dataset contains vintages for West Germany until November 1993. From February 1994 on-wards, the dataset provides vintages for the unified Germany. Unfortunately, the length of the vintages between February 1994 and August 1995 is critically short. Hence, we use vintages of GNP for West Germany until May 1995 (provided by \cite{boysen2012impact}). From August 1995 onwards, we use GDP for united Germany provided by the Federal Reserve of Dallas. Finally, starting from 1999, we use data on GDP provided by the OECD Main Economic Indicators database.\\

\subsection{VAR Data}
\underline{Quarterly GDP Japan}\\
\noindent We obtain quarterly real GDP for Japan from 1994Q1 until now from the OECD database. The Economic and Social Research Institute of Japan provides Real GDP Growth prior to 1994 at 2000 prices that is used to calculated from Real GDP until 1980. Real GDP Growth prior to Q1 1981 is calculated from the last growth rate of of each window for a certain quarter from our release dataset.

\subsection{Benchmark Revisions}\label{appendix: benchmark}
\noindent In this paper, we separate revisions of macroeconomic variables into revisions because of news and revisions because of noise. However, besides news and noise, there exists an additional source of revisions that we have not considered so far: benchmark or comprehensive revisions. While revisions because of news and noise represents revisions in the data that are because of new information or measurement errors, benchmark revisions represent revisions in the data that are because of changes in the GDP calculation process. These revisions can include the incorporation of new data sources as well as the adaptation of major changes in the statistical methodology. In addition, Benchmark revisions can be because of changes in the GDP definition. As the economy evolves over time, i.e. new sector appear, others disappear, the definition of GDP needs to be adopted to fit best the evolved structure of the economy. Benchmark revisions usually take place every four to five years and comprise comprehensive changes to the way GDP is calculates. Since World War II the were 15 benchmark revisions in the United States (see Table \ref{tab:benchmark_us}). \\

\begin{table}[!htbp]
 \caption{NIPA Comprehensive Revisions}\vspace{-0.75cm}
    \begin{center}
    \scalebox{0.68}{
    \begin{tabular}{llcll}
        Number       & Date  &   & Number  & Date \\
    \cellcolor{red_s} 1$^{st}$       & \cellcolor{red_s} July 1947  &\cellcolor{white} & \cellcolor{green_s} 9$^{th}$       & \cellcolor{green_s} December 1991  \\
    \cellcolor{red_s} 2$^{nd}$       & \cellcolor{red_s} July 1951  &\cellcolor{white} & \cellcolor{green_s} 10$^{th}$       & \cellcolor{green_s} January 1996\\
    \cellcolor{red_s} 3$^{rd}$       & \cellcolor{red_s} July 1954  &\cellcolor{white} & \cellcolor{green_s} 11$^{th}$       & \cellcolor{green_s} October 1999  \\
    \cellcolor{red_s} 4$^{th}$       & \cellcolor{red_s} July 1958  &\cellcolor{white} & \cellcolor{green_s} 12$^{th}$       & \cellcolor{green_s} December 2003 \\
    \cellcolor{red_s} 5$^{th}$       & \cellcolor{red_s} August 1965  &\cellcolor{white} & \cellcolor{green_s} 13$^{th}$       & \cellcolor{green_s} July 2009\\
    \cellcolor{green_s} 6$^{th}$       & \cellcolor{green_s} January 1976  &\cellcolor{white} & \cellcolor{green_s} 14$^{th}$       & \cellcolor{green_s} July 2013\\
    \cellcolor{green_s} 7$^{th}$       & \cellcolor{green_s} December 1980  &\cellcolor{white} &\cellcolor{green_s} 15$^{th}$       & \cellcolor{green_s} July 2018 \\
    \cellcolor{green_s} 8$^{th}$       & \cellcolor{green_s} December 1985  &\cellcolor{white} & &
    \end{tabular}
    }
    \end{center}
        \footnotesize{Notes: The table lists the dates of the 15$^{th}$ comprehensive revisions of the U.S. National Income and Product Accounts (NIPA) since World War II. The dating of the comprehensive revisions is taken from \cite{fox2017methods} for the 1$^{st}$ to the 14$^{th}$ revision. The date of the 15$^{th}$ comprehensive revisions is taken from the BEA website (https://apps.bea.gov/scb/2018/04-april/0418-preview-2018-comprehensive-nipa-update.htm). Green rows mark comprehensive revisions that are within the years covered by our uncertainty measure for the United States. Rows in red mark comprehensive revisions prior to the start of our uncertainty measure.}
    \label{tab:benchmark_us}
\end{table}

\begin{figure}[!htbp]
    \centering
    \scalebox{0.78}{
    \includegraphics{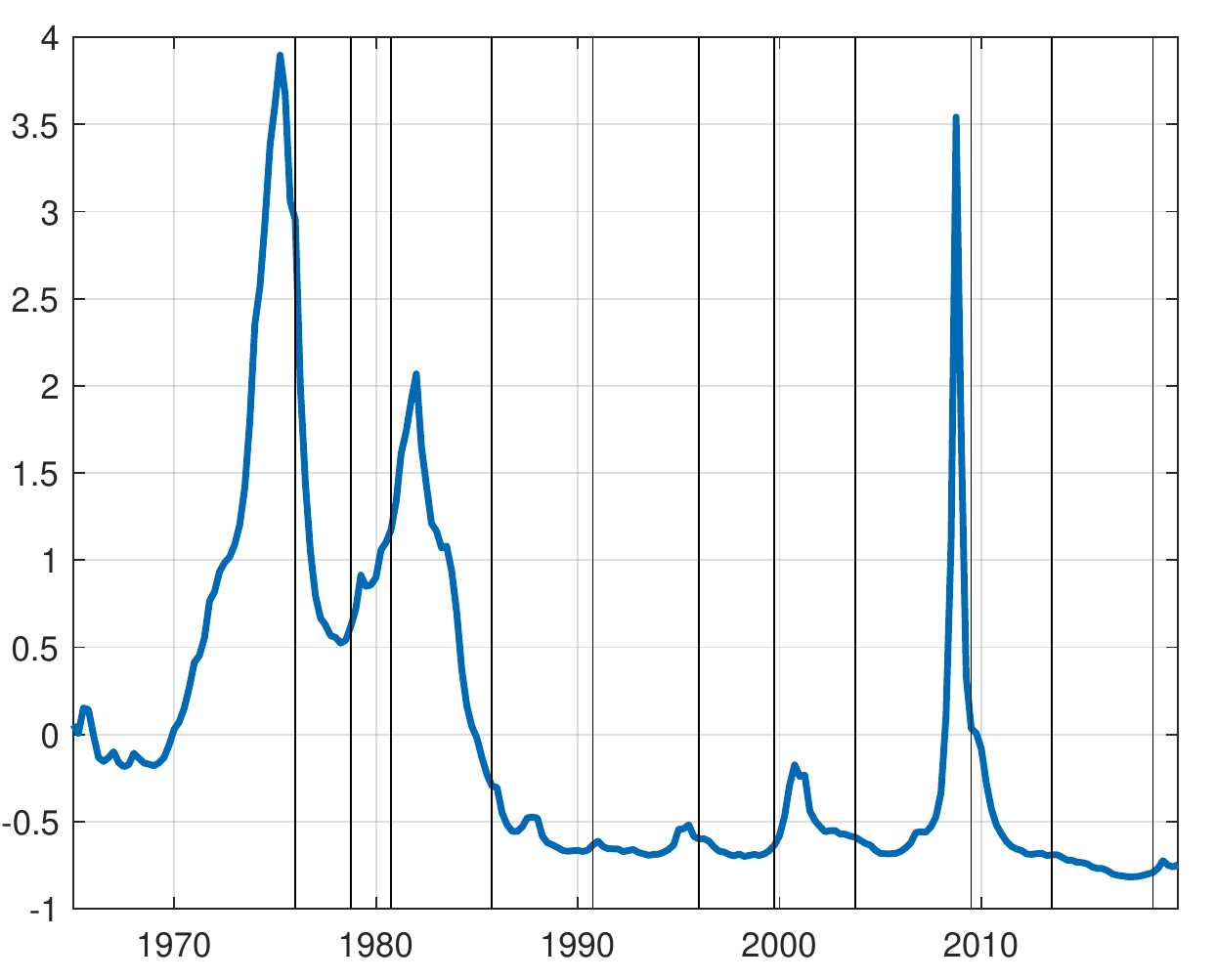}
    }
    \vspace{0cm}\caption{Impact of Benchmark Revisions \newline \scriptsize{\newline
						Notes: The figure presents the macroeconomic uncertainty indicator for the United States and highlights the dates of relevant benchmark revisions. The dates of the benchmark revisions correspond to the dates of Table \ref{tab:benchmark_us}. }
		}\label{fig:benchmark}
\end{figure}

\noindent Benchmark revisions can lead to large and significant revisions in the level of macroeconomic variables. Large revisions in GDP that are not because news or noise would of course be a concern for our estimation procedure. Luckily, while benchmark revisions can lead to significant changes in levels, growth rates of macroeconomic variables are much less effected by benchmark revisions. However, even though benchmark revisions are much less of a concern when working with growth rates, we still want to examine if benchmark revision have a systematic effect on our uncertainty measure. Figure \ref{fig:benchmark} plots our uncertainty indicator and highlights the various benchmark revisions. Overall, we have 10 benchmark revisions that occurred during the time period covered by our indicator. Figure \ref{fig:benchmark} does not expose any clear pattern of influence. While some benchmark revision occurred close to spikes of our indicator, others occur in periods where our indicator does not react at all. We take this as descriptive evidence that benchmark revisions are unlikely to systematically cause major spikes of our uncertainty indicator.

\newpage

\section{Uncertainty Estimates}\label{sec:app__unc}

\begin{figure}[!htb]
    \centering
    \scalebox{0.60}{
    \includegraphics{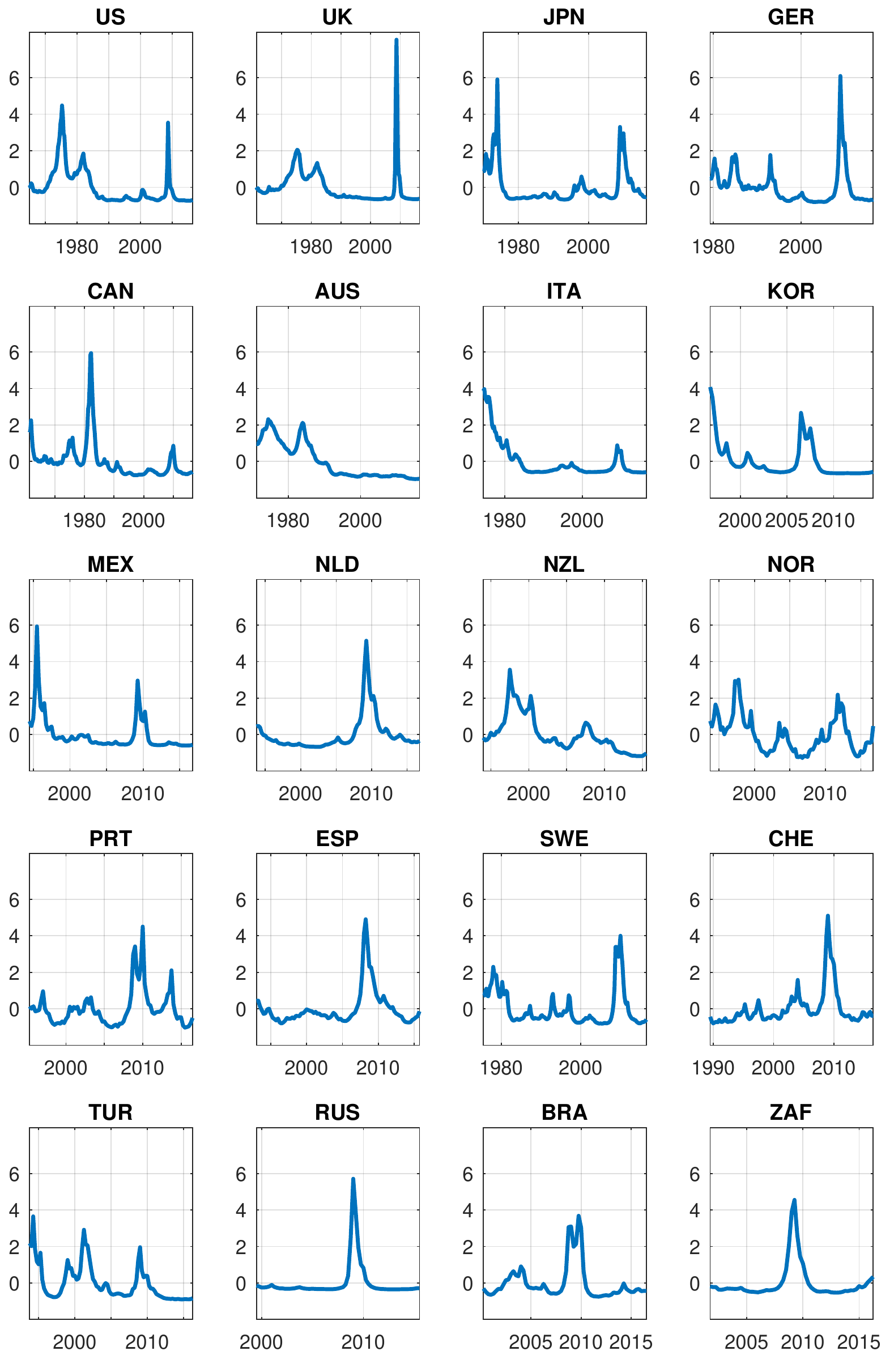}
    }
    \vspace{0cm}\caption{Macroeconomic Uncertainty \newline \scriptsize{\newline
						Notes: This figure presents uncertainty indicators based on real GDP growth revisions. All indicators are demeaned and standardized to unit variance.}
		}\label{fig:allUnc1}
\end{figure}

\begin{figure}[!htbp]
    \centering
    \scalebox{0.60}{
    \includegraphics{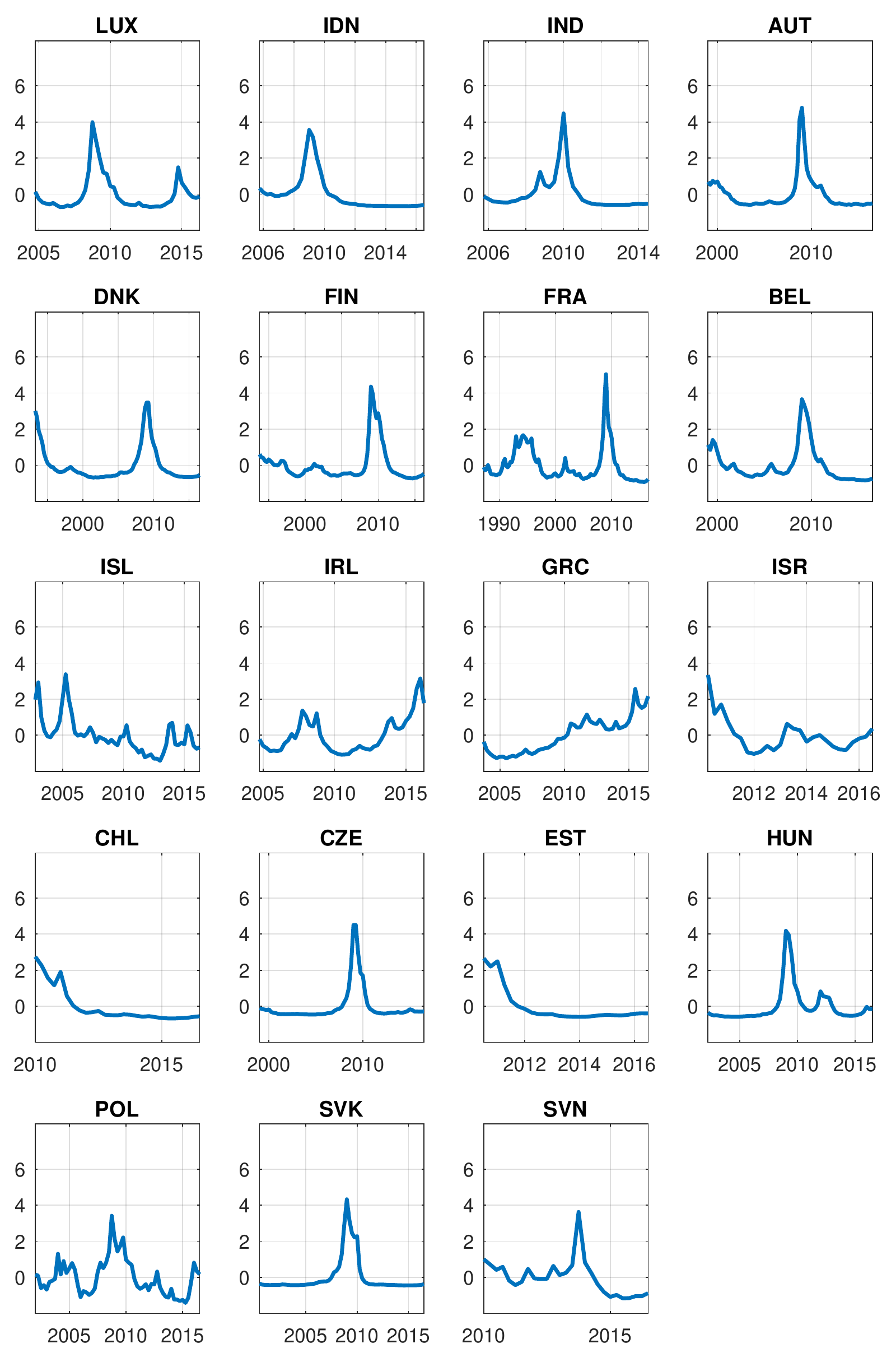}
    }
    \vspace{0cm}\caption{Macroeconomic Uncertainty (cont.) \newline \scriptsize{\newline
						Notes: This figure presents uncertainty indicators based on real GDP growth revisions. All indicator are demeaned and standardized to unit variance.}
		}\label{fig:allUnc3}
\end{figure}

\newpage

\section{Information Content of the First Release of GDP}\label{sec:app__gdp}

Statistical agencies publish the first GDP figure for a given quarter about one month after the end of a quarter. At that point in time, they do not have all information necessary to compute the GDP. Thus, some parts of GDP need to be estimated. While little is known about the exact estimation procedure applied by statistical agencies, \cite{scruton2018introducing} describe the GDP data collection process of the Office for National Statistics (ONS) in the United Kingdom. In their report, the authors describe the ONS GDP publication routine that was valid until 2018. The ONS publishes three GDP releases per quarter. The first figure of GDP is released about 25 days after the end of a quarter. At this point in time, ONS has only output data at their disposal. The available information content corresponds to about 45\%. The rest needs to be forecasted by ONS. \cite{scruton2018introducing} state that ONS forecasts large parts of the final month of a quarter as less data tend to be available for the last month of a quarter compared to the previous two months. Approximately two months after the end of a quarter, ONS publishes the second estimate of GDP. This estimate builds already on a significantly larger information set and includes, besides output data also data on income and expenditure.The overall data content amounts to approximately 65\%. ONS publishes the third estimate of GDP roughly 85 days after the end of a quarter and the overall data content contained in the estimate amounts to approximately 90\%.\\

\noindent The insights provided by \cite{scruton2018introducing} supports our view that one can interpret the first release of GDP as a forecast by statistical agencies for the final true GDP. In this paper, we treat the first published year-over-year growth rate of real GDP as a forecast for the final release of GDP, which we define as the year-over-year growth rate released after 3 years. This view allows us to compute quarterly forecast error made by the statistical agencies. We follow the idea by \cite{Juradoetal2015} and compute the conditional volatility of the forecast error series which we define as our measure of macroeconomic uncertainty. Thus, this paper presumes that GDP forms a good summary of the macroeconomic activity in an economy.\\

\noindent In their paper, \cite{Juradoetal2015} argue that it is important to remove the entire forecastable component of a forecast error in order to obtain the true unpredictable part of a forecast. Otherwise the explainable part will incorrectly end up in the uncertainty component. In order to obtain the unpredictable part of the GDP revisions (i.e., the unpredictable forecast errors committed by a statistical agent), we use the model outlined in Section \ref{sec:econometric}. Our measure of macroeconomic uncertainty is defined as the conditional volatility of the error corresponding to the unpredictable part of revisions in GDP growth rates.\\

\noindent Despite the decomposition of the revisions in into news and noise, it is key to understand the information content of the first release of GDP for the last release. In other words, we need to understand if the first release of GDP is a good forecast for the final GDP. Ideally, we want the first release to be the best available forecast for the final release of GDP. Given this central importance for our uncertainty measure, we test the information content of the first release of U.S. GDP for the last release of U.S. GDP by conducting a forecasting horse race.\\

\noindent In this forecasting horse race, we treat the first release itself as a forecast and the difference between the last release and the first release as forecast errors. Doing so allows us to compute the RMSE of these forecast errors and compare it to the following two pseudo out-of-sample forecast exercises for the final U.S. GDP release. Both forecasting exercises use the econometric framework outlined in Section 2 in \cite{Juradoetal2015} to forecast last release of GDP.\\

\noindent The first forecasting exercises uses the 132 macroeconomic variables and the 147 financial variables proposed by \cite{Juradoetal2015} to forecast the final GDP release. In order to conduct this forecast, we use the original dataset provided by the authors in their Online-Appendix. As the dataset contains monthly time series, but we need to forecast quarterly GDP, we aggregate the monthly series to quarterly series by taking quarterly averages. Real-time GDP for the U.S. starts in 1965Q1. The dataset by \cite{Juradoetal2015} provides data up to 2011Q4. We conduct pseudo out-of-sample\footnote{Unfortunately, we do not have real-time data of all series contained in the dataset by \cite{Juradoetal2015}. Using the provided \textit{last} vintages represents an information advantage. However, this information advantage works against the first releases.} forecasts for the quarter 1976Q4 to 2011Q4. In the second exercise, we repeat the first exercise, but use the least release as an additional predictor.\\ %We save the forecast errors for all three specifications and evaluate them using a Diebold-Mariano test \citep{diebold2002comparing}. 
%The Diebold-Mariano test reveals that using the 279 series does not improve final GDP forecasts relative to the first release. Moreover, adding the first release as an additional predictor to the model by \cite{Juradoetal2015} does not enhance forecasting performance.\\

\noindent This analysis shows that for the United States the first release of GDP appears to be a good forecast for the final GDP. We can show that when using established forecast routines and one of the most comprehensive dataset about the U.S. economy, we cannot beat the forecast performance of the first release.\\

\begin{table}[!htbp]
 \caption{Comparison Forecast Performance}\vspace{-0.75cm}
    \begin{center}
    \scalebox{0.68}{
    \begin{tabular}{lcccc}
          & First Release & JLN  & JLN    \\
          & (as forecast for last release) &  & +First Release    \\
          &  &  &     \\
    RSME  & 0.00752 &   0.00807   & 0.00816     \\
    Ratio (RMSE Only First Release) &   1.000 & 1.072 & 1.084

    \end{tabular}
    }
    \end{center}
       \footnotesize{Notes: The table compares the forecast accuracy in predicting the final year-over-year growth of real GDP in the United States. The first column refers to the forecast performance of the first release for the last release. The second column reports the performance of the forecasting model proposed by \cite{Juradoetal2015}. The last column refers to the model of \cite{Juradoetal2015} augmented with the first release as an additional predictor. In the first row, we display the RMSE error of the different specifications. The second row reports the RMSEs relative to the RMSE of the first release.}
    \label{tab:gdp}
\end{table}

\section{Robustness}\label{sec:app__rob}
\subsection{Selected High and Low EPL Countries}
\begin{figure}[!htb]
    \centering
    \includegraphics[scale=1]{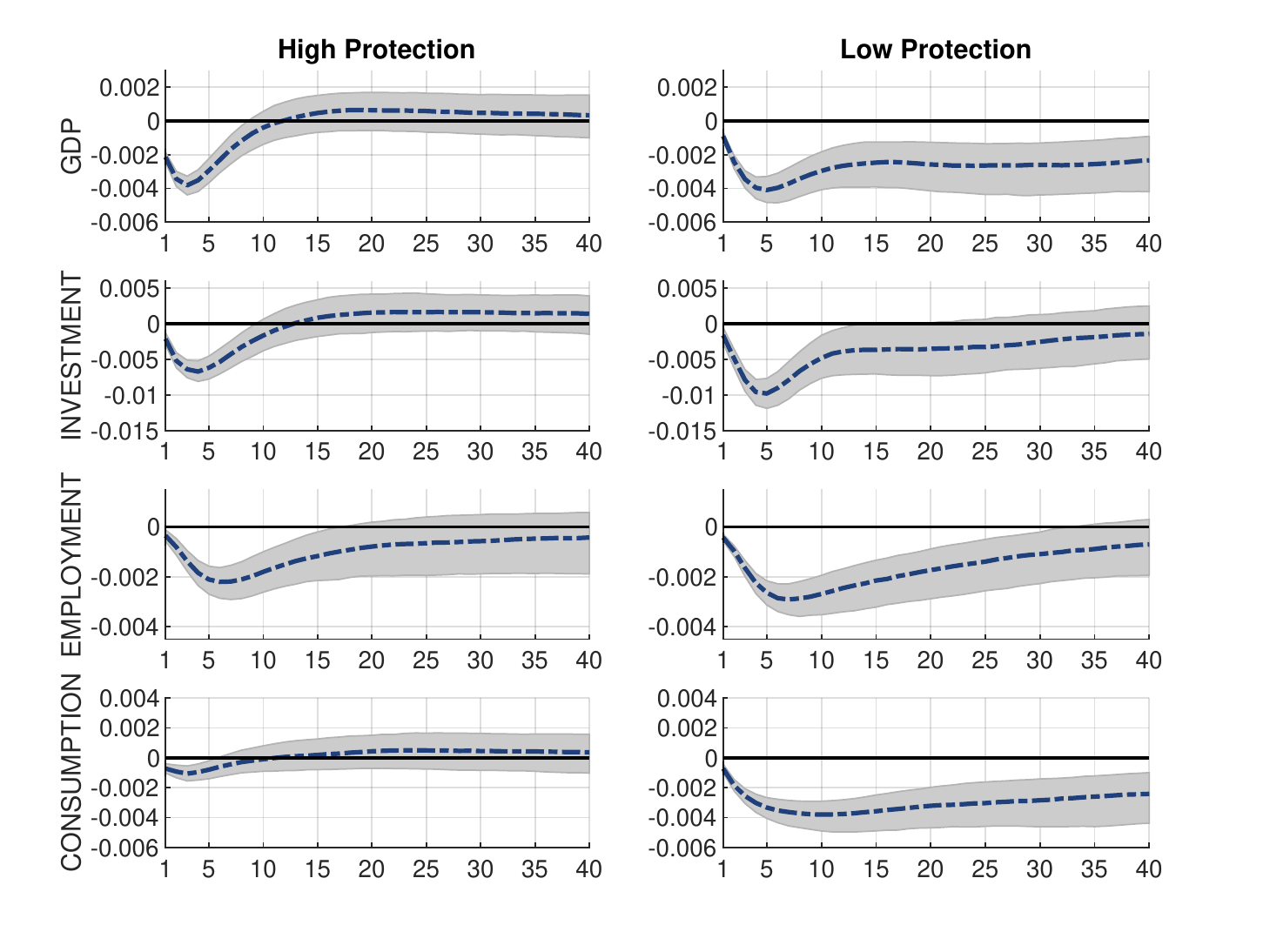}
            \caption{Impulse responses to an uncertainty shock for high EPL countries (left panel) and low EPL countries (right panel) without Japan and Switzerland. 
    \newline \scriptsize{\newline
			Notes: The dotted blue line depicts the posterior mean and the grey shaded area the 68\% error bands for the impulse responses to an one standard deviation uncertainty shock. The estimation sample spans the period 1988Q1--2016Q3.} }
    \label{fig:irf_hl_short}
\end{figure}

In order to empirically evaluate the effect of employment protection legislation (EPL) on the propagation of uncertainty shocks, we estimate a VAR for countries with stricter EPL and a VAR for countries with a lower degree of EPL. As mentioned in the main text, we use the OECD Employment Protection Database to split countries into two groups. The groups selected in Table \ref{t:epl} mirror our expectations with Anglo-Saxon economies displaying a low degree and continental European countries showing higher degree of employment protection standards. The OECD Employment Protection Database attests Switzerland and Japan have a very similar degree of employment protection legislation. In our baseline specification, we include Switzerland in the group with low labor protection and Japan in the group of high EPL countries. As a robustness test, we re-run the analysis excluding both Japan and Switzerland. Figure \ref{fig:irf_hl_short} shows that neglecting the two countries does not significantly change the results. Estimating a pooled VAR for both groups shows that uncertainty has still less deteriorating effects in countries with high employment protection.\\ 

\subsection{Incorporating Estimation Uncertainty}
\begin{figure}[!htb]
    \centering
    \includegraphics[scale=1]{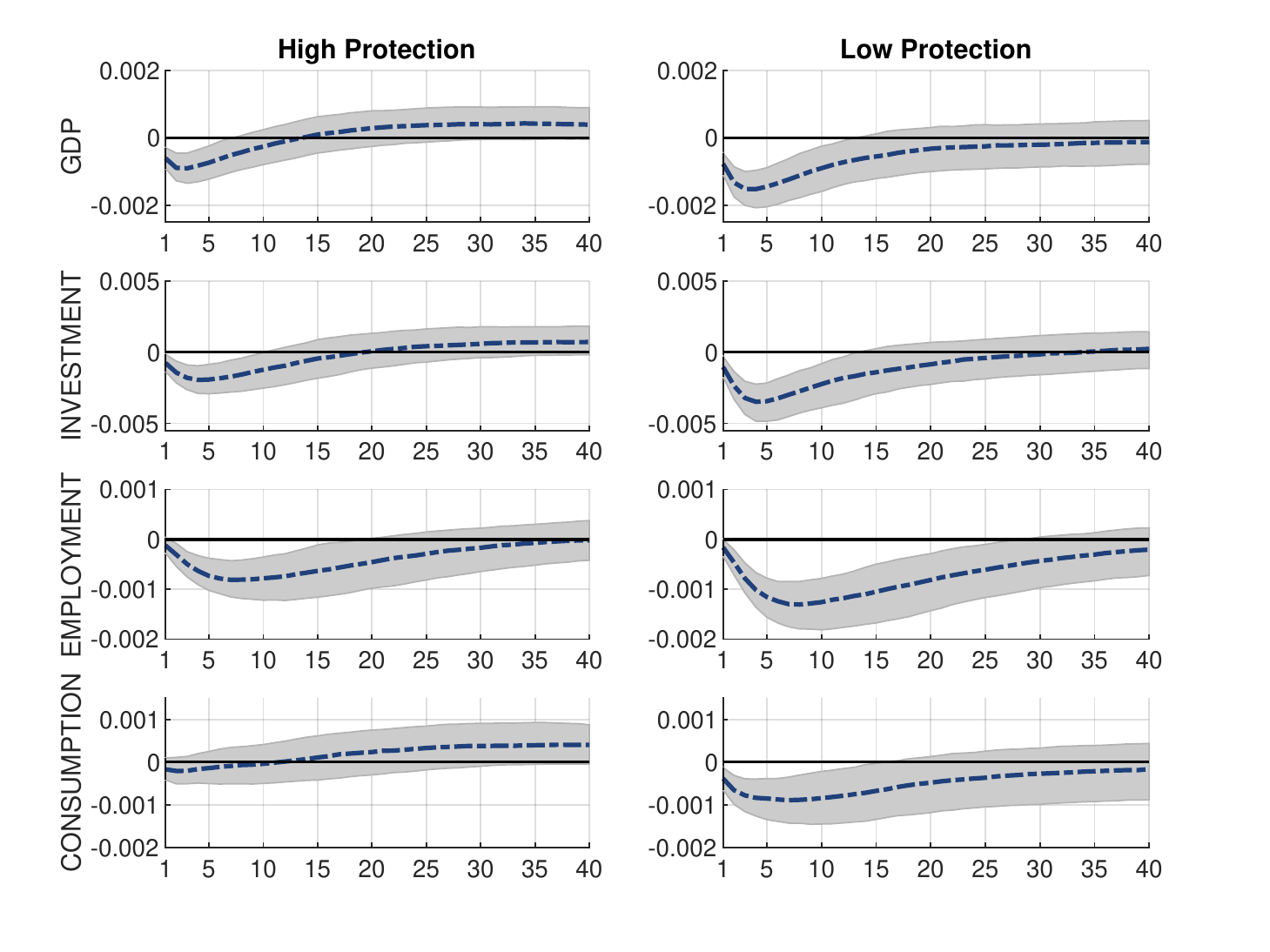}
            \caption{Impulse responses to an uncertainty shock for high EPL countries (left panel) and low EPL countries (right panel), including the uncertainty surrounding data revisions based uncertainty indicator. 
    \newline \scriptsize{\newline
			Notes: The dotted blue line depicts the posterior mean and the grey shaded area the 68\% error bands for the impulse responses to an one standard deviation uncertainty shock. The estimation sample spans the period 1988Q1--2016Q3.} }
    \label{fig:irf_hl_genreg}
\end{figure}

\begin{figure}[h]
    \centering
    \includegraphics[scale=0.8]{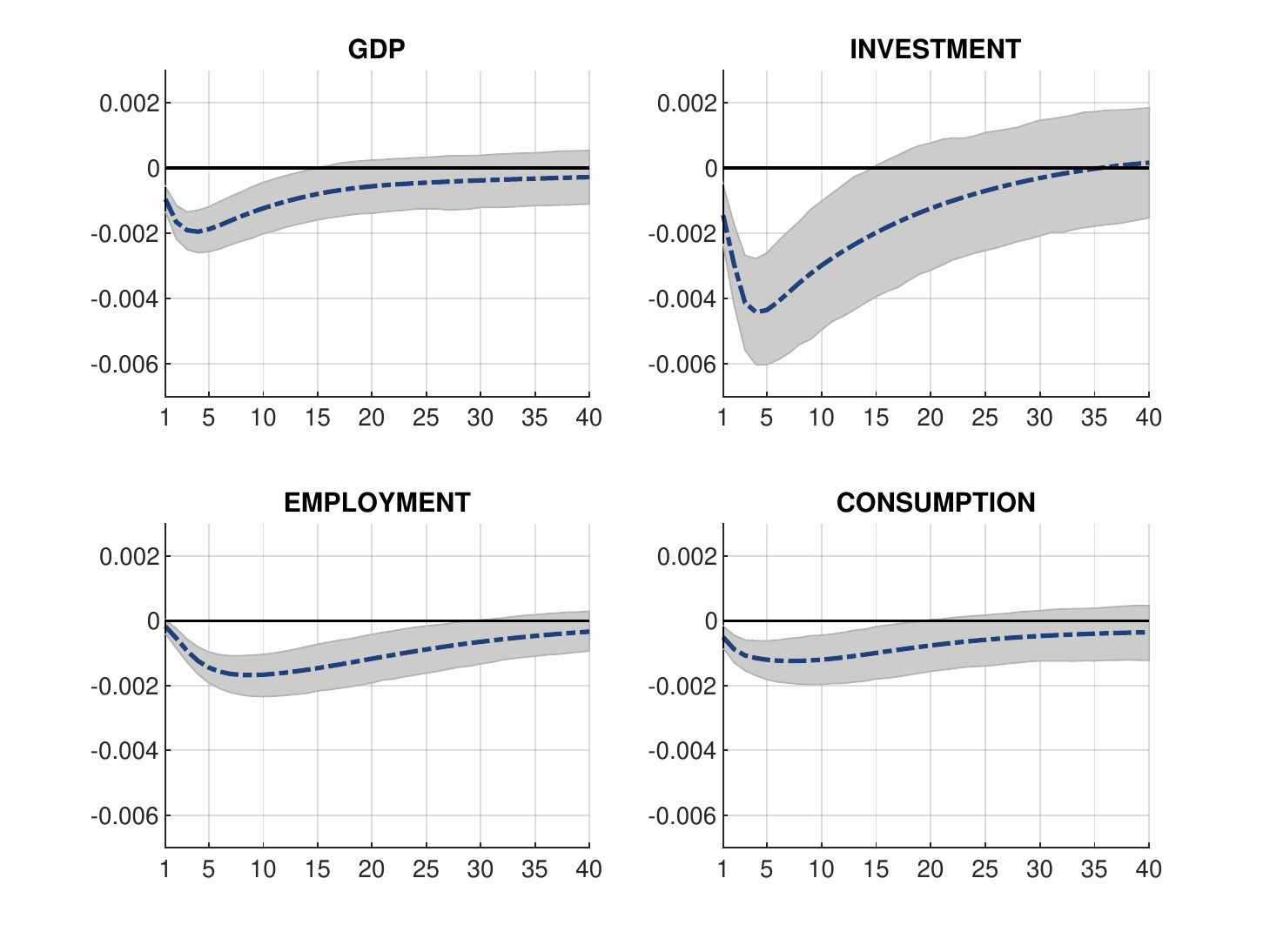}
    \caption{Impulse responses to an uncertainty shock for the group of G7 countries, including the uncertainty surrounding data revisions based uncertainty indicator.
    \newline \scriptsize{\newline
			Notes: The dotted blue line depicts the posterior mean and the grey shaded area the 68\% error bands for the impulse responses to an one standard deviation uncertainty shock. The estimation sample spans the period 1988Q1--2016Q3.} }
    \label{fig:irf_all_genreg}
\end{figure}

\noindent We now investigate the impact of the estimation uncertainty surrounding our measures of macroeconomic uncertainty on our main findings. Instead of using the mean of the posterior distribution of our uncertainty measure, we now simulate the posterior distribution of the VAR model conditional on the draws from the posterior distribution of our uncertainty indicator. Figure \ref{fig:irf_hl_genreg} compares the impulse responses functions of high protection countries to low protection countries. Although the overall effects are somewhat weaker than the ones shown in Figure \ref{fig:irf_hl}, the relative precision does not change in a significant manner. Most importantly, our principal findings do not change when taking into account the uncertainty surrounding our uncertainty estimates. Figure \ref{fig:irf_all_genreg} presents the overall effects for the G7 average. The responses of the variables seem also somewhat weaker than the ones reported in Figure \ref{fig:irf_all}. The impulse response functions of consumption are additionally not clearly negative anymore. The impulse responses for all other variable however appear robust with regards to the inclusion of the added estimation uncertainty. 

\end{appendices}

\end{document}